\documentclass[twocolumn,
 amsmath,amssymb,
 aps,
pra,
10pt,
]{revtex4-2}

\newcommand*\mathinhead[2]{\texorpdfstring{$#1$}{#2}}
\usepackage{array}[=2016-10-06]
\usepackage{placeins}
\usepackage{float}

\newcolumntype{C}[1]{>{\centering\let\newline\\\arraybackslash\hspace{0pt}}m{#1}}
\usepackage[titletoc,title]{appendix}

\usepackage[utf8]{inputenc}

\usepackage{graphicx}
\usepackage[shortlabels]{enumitem}
\usepackage{siunitx}
\setcitestyle{square}
\usepackage{enumitem}
\setlist[enumerate]{itemsep=0mm}
\usepackage{hyperref}
\hypersetup{colorlinks=true, citecolor=blue, urlcolor=blue, linkcolor=blue}
\usepackage{fancyhdr}
\usepackage{wrapfig}

\usepackage{verbatim}
\usepackage[bottom]{footmisc}

\begin{document}

\title{\textbf{A Penning trap single-photon counter for axion detection} 
}%

\author{Jack A. Devlin}%
\email{Corresponding author: j.devlin11@imperial.ac.uk}
\author{Marko L. Wojtkowiak}%
\author{Shreyak R. Banhatti}%
\author{He Zhang}%
\author{Jiacheng Shi}%
\author{Toren S. Dofher}%
\author{Jonathan M. H. Gosling}%
\author{Michael R. Tarbutt}%
\author{Richard C. Thompson}%
\affiliation{ Centre for Cold Matter, Blackett Laboratory, Imperial College London,
Prince Consort Road, London SW7 2AZ, United Kingdom
}%

\date{\today}


\begin{abstract}
Discovering the microscopic composition of dark matter is one of the most important open problems in physics today. Axions are a leading candidate to be dark matter; however, a search of the full range of all likely axion masses is hampered by the standard quantum noise limit. This makes haloscope searches for axions with masses above ${\sim}100$~\si{\micro\electronvolt} unfeasible with current technologies. To overcome this limitation, we propose a new photon counting technique designed to operate at {30--60~GHz} for detecting axions with masses between 124 \si{\micro\electronvolt} and 248 \si{\micro\electronvolt}, based on a single electron in a Penning trap. The electron cyclotron mode absorbs microwave photons, and, via the continuous Stern-Gerlach effect, this absorption imparts a measurable phase shift onto the axial motion. In this paper, we comprehensively analyze this photon detection method. We introduce a new type of fast, phase-sensitive axial detection technique, using axial-magnetron parametric amplification to overcome detector Johnson noise and cancel associated frequency shifts. This method may find other applications in precision Penning trap frequency measurements. We compare the efficiency of the electron single-photon counter with an ideal device, and find that our proposed photon counter has sufficient performance to search for high mass axions. 
\end{abstract}

\maketitle

\section{Introduction}
Dark matter is an unknown substance which affects astrophysical observations on sub-galactic to cosmological distances. There is no Standard Model particle which fits the bill to be dark matter, and considerable efforts are underway to experimentally discover new particles that could be this missing mass. One plausible particle that could be dark matter is the axion. Axions are pseudoscalar bosons, first proposed to explain why CP symmetry appears to be conserved by the Strong force. Axions are predicted not to be completely dark, with an axion-to-photon interaction Lagrangian, in natural units, of 
\begin{align}
\mathcal{L}_a=-\frac{1}{4}g_{a\gamma\gamma}aF^{\mu\nu}\tilde{F}_{\mu\nu},\label{eq:axionPhotonCoupling}
\end{align}
where $F$ is the electromagnetic tensor, $\tilde{F
}$ is its dual, $a$ is the axion field and the strength of the interaction is governed by the axion-to-photon coupling constant, again in natural units
\begin{align}
    g_{a\gamma \gamma}=C_{a\gamma }\frac{\alpha}{2 \pi }\frac{m_a}{\sqrt{\chi}}.
\end{align}
Here $\alpha$ is the fine structure constant, $m_a$ is the axion mass and $\chi=[75.44(34) \textrm{MeV}]^4$ is a QCD parameter called the topological susceptibility \cite{Gorghetto2019TopologicalCorrections} and $C_{a\gamma}$ is a dimensionless coupling constant. There are various models which produce axions that solve the Strong CP problem, and the choice of model affects the value of $C_{a\gamma}$, sometimes also written as $g_\gamma=C_{a\gamma}/2$. Two well established models are the KSVZ \cite{PhysRevLett.43.103, SHIFMAN1980493} and DFSZ \cite{DINE1981199,zhitnitskij_1980} models, for which $C_{a\gamma}=-1.92(4)$ and $C_{a\gamma}=0.75(4)$ respectively \cite{Navas2024ReviewPhysics}. Thus, although the axion mass is not predicted, there is a well understood relation between the axion mass and the coupling strength. Experiments have ruled out axions with masses above 1 eV, but sub-eV axions remain a viable and well-motivated dark matter candidate, and there are production mechanisms in the early universe which straightforwardly generate axion abundances consistent with the $\Lambda$CDM model and observed cold dark matter density \cite{PRESKILL1983127, ABBOTT1983133, DINE1983137, Marsh2016AxionCosmologyb}. In our universe, axions would form a gravitationally bound dark matter halo, which could interact with a terrestrial ``haloscope'' detector. Most experiments seek to use the axion-photon coupling given by Eq.~\ref{eq:axionPhotonCoupling} to convert axions into rf photons. Building a device with sufficient sensitivity to produce a detectable signal from axion-normal matter interactions is challenging. Single-photon counters dramatically improve the search for axions with masses above about ${\sim}100$ \si{\micro\electronvolt}, potentially boosting the rate at which different axion masses can be investigated by many orders of magnitude. This application requires low dark counts and high duty cycle; currently, there are no single-photon counters with the necessary characteristics in this frequency range. In this work, we propose that an electron in a Penning trap device could fill this niche.    

Trapped electrons have been used for a rich series of fundamental physics tests. They can be used for measurements of the electron $g$-factor \cite{Hanneke2008NewConstant} and leptonic CPT tests \cite{VanDyck1987NewFactors} as well as quantum computing in both Paul \cite{Yu2022FeasibilityElectrons, Peng2017SpinQubits, Daniilidis2013QuantumElectronics, Matthiesen2021TrappingTrap, PhysRevResearch.4.033245} and Penning traps  \cite{Bushev2008ElectronsProcessing,Marzoli2009ExperimentalComputer, Goldman2010OptimizedStudies}. Trapped electrons have also been proposed as sensors for directly detecting dark matter-normal matter collisions \cite{ PRXQuantum.3.010330,Carney2021TrappedDetectors} and the conversion of dark photons to normal photons \cite{Fan2022One-ElectronDetector}, part a range of new experiments to detect dark matter with quantum techniques \cite{Devlin01102024}. Electrons can also be used to detect very weak electromagnetic radiation. Individual thermal microwave photons inside a cavity have been detected using electrons in a Penning trap \cite{Peil1999ObservingStates}. The microwave photons were observed to drive transitions between the modified cyclotron states of the trapped electron, and these transitions could be detected by measuring the electron axial frequency. In this paper, we propose an extension of this technique to detect incoming microwave photons produced outside of the trap, so that a trapped electron can serve as a single-photon counter. The basic physics behind our concept is similar to proposals by Cridland et al. \cite{Cridland2016SingleElectron}, and Fan et al. \cite{Fan2025HighlyDetection}, however, we choose a different experimental configuration and measurement protocol to achieve photon detection.  

The structure of the paper is as follows: in Section \ref{sec:2}, we consider why single-photon counters are vital for certain experiments searching for dark matter axions, and we review the technologies for single microwave photon counting at the relevant microwave frequencies. In Section \ref{sec:overview}, we give an overview of the method for counting microwave photons with a trapped electron. Section \ref{sec:4} deals with the cavity QED interaction between the electron and the cavity, and how this leads to an additional phase shift on the electron's axial motion, conditional on a photon being absorbed by the electron. In Section \ref{sec:5}, we consider how the axial and magnetron modes evolve during the photon counting sequence, and describe the noise sources that enter into a measurement of the electron axial phase, which affect our ability to resolve the signal from a photon absorption. Section \ref{sec:6} presents the detection efficiencies expected under various experimental configurations. Finally, in Section \ref{sec:7} we discuss our results, compare these to other photon counting proposals at these frequencies, and suggest avenues for further improvements in efficiency.      

\section{Single-photon counters for axion search experiments}
\label{sec:2}
In a typical haloscope experiment to detect axions, a resonant cavity is placed in a strong magnetic field, and the output of a particular mode of the cavity is monitored. If the axion (angular) Compton frequency $\omega_a=m_a(c^2+v^2/2)/\hbar$, found from the axion mass $m_a$ and speed $v$ matches the haloscope cavity's resonant frequency, then some axions may be converted into photons inside the cavity at a frequency $\omega_a$. The nonzero range of axion speeds leads the converted rf power to have a frequency width $\Delta\omega_a=\omega_a/Q_a$, where the axion quality factor $Q_a\sim10^6$, independent of the axion mass. For higher axion masses, there are also other approaches, including using magnetized mirrors \cite{PhysRevLett.128.131801,Brun2019AmathrmeV} or using plasma resonances \cite{PhysRevD.107.055013}, but the signal in all cases is the same: an expected increase in the number of photons emitted from the structure, when it is configured to be sensitive to a particular range of axion frequencies.  

The axion-sourced rf power coupled out of a haloscope is given by \cite{Kim2020RevisitingHaloscopes, Graham2024Rydberg-atom-basedSearches} 
\begin{align}
    P_a&=\left(\frac{\alpha^2\hbar^3c^3 C_{a\gamma}^2\rho_{DM}}{4\pi^2\mu_0 \chi}\right)\left(\kappa_c \omega_a B^2V_m\frac{Q_cQ_a}{Q_c+Q_a}\right) .
\end{align}
The factors inside the first pair of brackets are out of the experimenter's control. In addition to the factors introduced previously, there is also $\rho_{DM}$, the local dark matter density. In the second bracket are factors within experimental control (apart from $Q_a$): the coupling constant $\kappa_c$, which gives the ratio of power coupled out of the cavity to the total power lost per cycle, the loaded cavity $Q$-factor $Q_c$, the external magnetic field $B$ into which the cavity is placed, and the volume factor 
\begin{align}
    V_m=\frac{|\int E_m\cdot B dV|^2}{\int \epsilon_r|E_m\cdot B|^2 dV}.
\end{align} 
Using $\rho_{DM}=0.45\,\textrm{GeV/cm}^3$ and a critically coupled cavity of $\kappa_c=1/2$, the rate at which photons are emitted (in photons per second) is $R_a=\frac{P_a}{\hbar\omega_a}\simeq 5\times 10^{-4}C_{a\gamma}^2B^2V_mQ_h$, with all quantities in SI units and $Q_h=\frac{Q_cQ_a}{Q_c+Q_a}$. Typically, for $\omega_a=2\pi\times30$~GHz we expect $R_a$ to be around 1 count/s to 1 count/hour. 

Any experiment that searches for axions by converting them into microwave photons has to somehow detect the power emitted from the axion-to-photon converter. Traditionally, this has involved using a linear amplifier based on a HEMET, SQUID, or Josephson Parametric amplifier to increase the low-power microwave signal from axion decay. The amplified signal can then be down-mixed to a convenient frequency, digitized, and Fourier transformed in a spectrum analyzer. The axion signal would lead to excess noise at $\omega_a$, above the amplifier noise floor. To assess the size of a resolvable axion signal, the axion-sourced power needs to be compared to the noise at this particular frequency. For linear amplification, there is a fundamental lower limit to the noise floor that can be reached, even as both the cavity and detector approach zero temperature. This is referred to as the Standard Quantum Limit (SQL), and arises because linear amplification allows both the phase and amplitude of the field to be measured simultaneously, but these quantities obey a Heisenberg uncertainty relation which prevents simultaneous measurement at arbitrary precision \cite{Lamoreaux2013AnalysisSearches}. 

The SQL can be avoided by counting the photons, a process that does not preserve phase information and so is not subject to SQL noise. To compare single-photon counting to linear amplification for axion searches, the best figure of merit is the time required to scan a frequency range $\Delta\omega_a$ to the level where both a single-photon counter and a linear amplifier can detect or exclude an axion-sourced signal at the same level of significance. The ratio between the scan rates for single-photon counting and linear amplification is approximately given by  \cite{Graham2024Rydberg-atom-basedSearches} 
\begin{align}
    E&=\frac{Q_c}{2\pi Q_a}\frac{\eta_\textrm{sp}^2}{\eta_\textrm{la}^2}\left(\frac{\Delta\omega_a}{\eta_\textrm{sp}R_a+\eta_\textrm{sp}R_\textrm{T}+R_\textrm{RO}}\right) \,.\label{eq:enhancement}
\end{align}
Here $\eta_\textrm{sp}$ and $\eta_\textrm{la}$ are the detection efficiencies for single-photon counting and linear amplification, respectively, $R_\textrm{T}$ is the rate of thermal black body photons and $R_\textrm{RO}$ represents other non-thermal dark counts from the device readout. As $\Delta\omega_a=\omega_a/Q_a$, where $Q_a$ is independent of $\omega_a$, this expression suggests that single-photon counting will be more advantageous at higher axion frequencies. We can use this expression to consider the point at which photon counting is beneficial to an axion experiment. At low enough temperatures $R_\textrm{T}\ll R_a$. We also can typically control $R_\textrm{RO}$ to be much smaller than $\eta R_a$. For most current experiments, $Q_c/ Q_a$ is around 1/100, but in future cavity experiments, including those under development in our laboratory, it is feasible to imagine $Q_c/ Q_a\sim 1$. In this case, assuming $\eta_\textrm{la}=1$, the enhancement is
\begin{align}
    E&=\frac{\omega_a \eta_{\textrm{sp}}}{2\pi Q_aR_a}\nonumber\\&=3\times10^4\left(\frac{\eta_\textrm{sp}}{1}\right)\left(\frac{\omega_a}{2\pi\times30~ \textrm{GHz}}\right)\left(\frac{1~\textrm{cps}}{R_a}\right).
\end{align}
At 30 GHz, a single-photon counter with an efficiency of $\eta_\textrm{sp}>3\times10^{-5}\frac{R_a}{1~ \textrm{cps}}$ will outperform a linear amplifier. Typical count rates $R_a$ for accessible values of $B$, $V_m$ and $Q_h$ are of order 0.001-1 counts/s, suggesting huge boosts for even very low single-photon counter efficiencies. Assuming $R_a=0.1$ counts/s, a single-photon counter with $\eta_\textrm{sp}=0.5\%$ will outperform a linear amplifier by a factor of 1500 at 30 GHz, while $\eta_\textrm{sp}=10\%$ improves the scan rate by a factor 30,000 at the same frequency.   

The benefits of single-photon counting for axion experiments operating at frequencies above around 10 GHz have long been recognized \cite{Lamoreaux2013AnalysisSearches}. Unfortunately, the technology to count single photons in the window 30-300 GHz at high efficiency and low dark count rates does not currently exist. At frequencies above this band, transition-edge sensors \cite{Fukuda2024SingleSensors} and superconducting nanowires \cite{Yin2012SuperconductingApplications} typically operate at THz frequencies, whereas superconducting hot electron bolometers operate above 300 GHz. Work is underway to make transition edge sensors sensitive down to 90 GHz \cite{Alesini2020StatusDetection} to partially bridge this gap. Below this range of frequencies, circuit quantum electrodynamics (cQED), superconducting qubits interacting with microwave cavities, can be used to count single photons \cite{Blais2021CircuitElectrodynamics}. In the standard cQED configuration, these devices may not have dark count rates low enough for axion searches; however, recent progress \cite{Dixit2021SearchingQubit} using improved measurement protocols has demonstrated counting efficiencies up to 0.409(55) with false positive rates of $4.3(1.1)\times10^{-4}$ at 6.011 GHz with a duty cycle of 65\%. This method seems directly extendable up to 30~GHz or above, and recent progress has been made in this direction \cite{Anferov2024SuperconductingMK}. In the future, higher microwave frequencies could potentially be detected using superconducting qubits made from higher $T_c$ superconductors. Another approach is to use Al-based tunnel Josephson junctions in a current-biased regime \cite{Pankratov2022TowardsAxions}. If the Josephson junction is coupled to a photon field, it switches to a resistive state, and this leads to a voltage across the junction. Few-photon detection sensitivity was demonstrated at 10 GHz via this method. Single-photon counters are also proposed in the 30-300 GHz range using quantum dots \cite{Ghirri2020MicrowaveDots} and optomechanical sensors \cite{Zhang2015ProposalLevel}. Rydberg atoms have also been used in single-photon counting experiments with a view to conducting axion searches \cite{Tada1999CARRACKDetector,Tada2006Single-photonAtoms}, and are also proposed to explore the range 10-50 GHz \cite{Graham2024Rydberg-atom-basedSearches}. Finally, electrons in surface \cite{Cridland2016SingleElectron} and macroscopic \cite{Fan2025HighlyDetection} Penning traps have also been proposed to detect microwave photons. In the sections that follow, we present our method for counting single photons with a trapped electron. In Section \ref{sec:7} we contrast our approach with other methods and draw some conclusions.

\section{Overview of the photon detection method}
\label{sec:overview}
\begin{figure*}[t]
\centering
\includegraphics[]{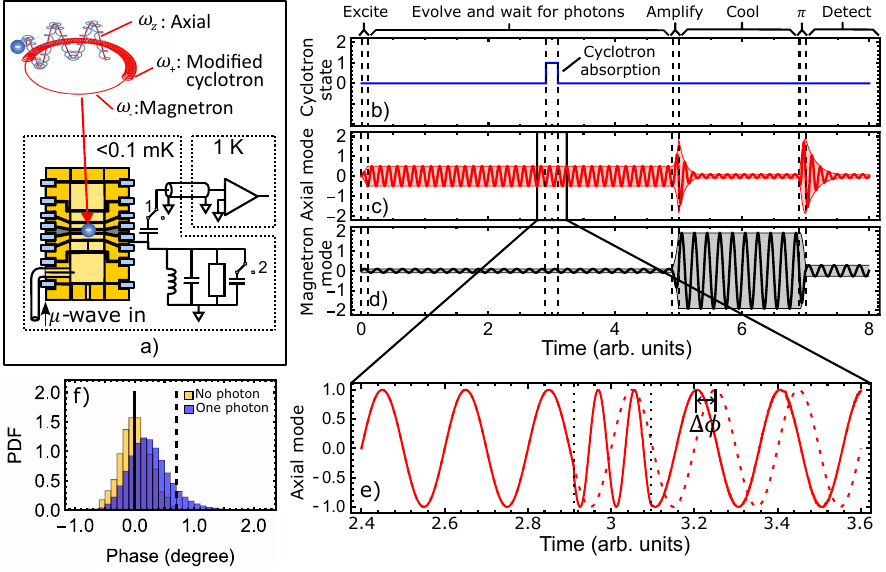}
\caption{An illustration of the photon counter and detection sequence. a) Shows the three motions in the Penning trap, and a cross-section of the trap and electronic detection circuit. b) Shows the cyclotron state during a photon detection sequence, c) illustrates the axial amplitude and d) shows the magnetron amplitude. e) An expanded version of the axial oscillation illustrating how a phase shift due to the cyclotron absorption event arises from a transient frequency shift. The size of the frequency shift has been exaggerated to make it more visible in this illustration. f) An illustration of a phase distribution after repeated photon counting sequences deliberately exaggerated to illustrate the effect: yellow gives the distribution when no photon was absorbed, and blue gives the distribution after an absorption.}
\label{figure:Plots}
\end{figure*}

Our microwave single-photon counter consists of a single electron in a cryogenic Penning trap.  The Penning trap, pictured in Fig.~\ref{figure:Plots}~a) confines the electron with mass $m$ in a strong static magnetic field $B_0 \boldsymbol{\hat{z}}$, and a quadratic electrostatic potential $\varphi=\frac{V_0 C_2}{2d^2}(z^2-\rho^2/2)$ generated by applying voltages to a closed set of cylindrical electrodes surrounding the electron. Here, the total effect of applying different voltages to the various electrodes of potentially different geometries is captured by a single effective voltage $V_0$, a characteristic length $d^2=\frac{1}{2}(z_0^2+\rho_0^2/2)$, determined by the trap radius $\rho_0$ and the distance between the trap center and the endcap electrode $z_0$, and a dimensionless constant $C_2$. In a hyperbolic trap, $V_0$ would be the real potential difference between the ring and endcaps. Neglecting relativistic effects, the Hamiltonian of this ideal system can be written as the sum of three quantum harmonic oscillators and a spin in a magnetic field \cite{Brown1986GeoniumTrap}
\begin{eqnarray*}
\begin{aligned}
    H_0=&\hbar\omega_+(a_+^\dagger a_++\tfrac{1}{2})+\hbar\omega_z(a_z^\dagger a_z+\tfrac{1}{2})\\&-\hbar\omega_-(a_-^\dagger a_-+\tfrac{1}{2}) +\frac{1}{2}\hbar\omega_s\sigma_z \, .
\end{aligned}
\end{eqnarray*}

Here ``+" denotes the modified cyclotron mode, ``$z$" the axial mode, ``$-$" the magnetron mode, $\sigma_z$ is the Pauli $z$ matrix, $a_i^\dagger$ the creation, and $a_i$ the annihilation operators for the respective modes. The mode oscillation frequencies for a trapped electron are given by $\omega_z^2=\tfrac{eV_0C_2}{m d^2}$, $\omega_\pm=\tfrac{1}{2}(\omega_c\pm(\omega_c^2-2\omega_z^2)^{1/2})$ and $\omega_s=\frac{g_S}{2}\omega_c$ where $\omega_c=\tfrac{e B_0}{m}$ is the free cyclotron frequency,  $g_S$ is the electron $g$-factor and $\omega_s$ is the Larmor frequency. The electron motion is illustrated in the top part of Fig.~\ref{figure:Plots}~a). By tuning the magnetic field $B_0$, the modified cyclotron frequency can be matched to the frequency $\omega_a$ of the microwave photon; to match photon frequencies between 30-60 GHz requires fields 1-2 T. In principle, higher frequencies could also be accessible with this method. The axial frequency, set by a choice of $V_0$ and the size of $d$, can be chosen to be in the range $\omega_z=2\pi\times 10-100$~MHz, with a corresponding magnetron frequency $\omega_-$ of a few tens to hundreds of kilohertz. The trap is held at temperatures $T_\textrm{e}<100$ mK by thermal connections to the mixing chamber of a dilution fridge. At this temperature, the cyclotron mode cools to its ground state via radiative emission with a time constant of $<1$ ms. The axial and magnetron modes can be brought to the same mode occupation number by coupling them using a rf drive at $\omega_z+\omega_-$, and then both modes can be cooled to the ground state by alternating this drive with a sideband coupling between the cyclotron and axial modes using microwaves at $\omega_+-\omega_z$. A resonant LCR circuit can be used to interact with the axial mode, as shown schematically in Fig.~\ref{figure:Plots}~a). It has three operating configurations depending on the position of the two switches shown. In the detection mode, the LCR circuit is resonant with the axial mode, and the switch labeled ``1'', which connects the LCR circuit via a capacitor to a transmission line and amplifier, is closed and the switch labeled ``2'' is open. This enables the amplitude and phase of the axial mode to be read out; however, the electronic temperature of the detection circuit in this configuration is rather high, $T\simeq1$~K, and the damping time constant is low, limited by losses in the amplifier, which reduces the $Q$-factor of the LCR circuit. The second configuration is the cooling configuration, in which switch 1 between the amplifier and the LCR coil is open. This allows rapid cooling of the axial mode to $T_\textrm{e}$ in situations where sideband cooling causes unwanted phase shifts. Finally, in the decoupled state, switch 1 remains open, and an additional capacitance is connected across the LCR circuit using switch 2 to shift it out of resonance with the axial mode, preventing both heating and damping of this mode. This is the default state of the resonant circuit.   

In our Penning trap, the electrodes surrounding the electron are designed to act as a resonant cavity, with a quality factor $Q$ and resonant frequency $\omega$, which is also adjusted to match $\omega_+$ and $\omega_a$, by control of the cavity endcap positions. Photons enter the cavity via a transmission line shown in the bottom left of Fig.~\ref{figure:Plots}~a) and once in the cavity, they can be absorbed by the cyclotron mode of the trapped electron. Adding an extra magnetic field of the form $\boldsymbol{ \Delta B}=B_2((z^2-\rho^2/2)\boldsymbol{\hat{z}}-z\rho\boldsymbol{\hat{\rho}})$ by means of a ferromagnetic ring electrode, pictured in gray in {Fig.~\ref{figure:Plots}~a)}, allows the absorption of the microwave photon to affect the axial frequency \cite{Brown1986GeoniumTrap}. This is because the magnetic field leads to an additional term in the Hamiltonian
\begin{align}
    H_1&=-\boldsymbol{\mu}\cdot\boldsymbol{\Delta B}.
\end{align}
The cyclotron mode has a magnetic moment $\boldsymbol{\mu}$ parallel to $\boldsymbol{\hat{z}}$, and the magnitude of the magnetic moment is proportional to the cyclotron quantum number $n_+$. A change in the cyclotron state following a photon absorption leads to a change in the magnetic moment equal to $|\Delta\boldsymbol{\mu}|$. For a particle oscillating along the $z$-axis the change in potential is therefore $|\Delta\boldsymbol{\mu}|B_2z^2$ imparting a frequency shift  
\begin{align}
    \Delta \omega _z&=\frac{\Delta n_+   \omega _+ \hbar B_2}{m \omega _z B_0} \label{EqAxialShift},
\end{align}
 to the axial frequency, where $\Delta n_+$ is the change in cyclotron quantum number.
 
Figs.~\ref{figure:Plots}~b)-e) illustrate the basic sequence of how a single-photon counting measurement is performed in our scheme, allowing the phase shift $\Delta\omega_z$ to be detected. The top graph, b), shows the cyclotron state of the particle. The particle begins in the cyclotron ground state. At $t=2.91$ the cyclotron state is briefly excited to $n_+=1$ for a time $t_\textrm{ex}$ by an incoming photon, before returning to the ground state. Fig.~\ref{figure:Plots}~c) shows the axial motion during the measurement sequence, while graph d) shows the magnetron motion projected onto one axis. Initially, during the period marked ``Excite'', the axial amplitude is excited with an rf pulse at a frequency $\omega_z$. There then follows a period labeled ``Evolve and wait for photon'', during which the axial motion freely evolves, and the experiment is sensitive to incoming photons. The absorption of the microwave photon causes a brief period where the axial frequency shifts, as described by Eq.~\ref{EqAxialShift}. This is shown more clearly in the magnified graph, Fig.~\ref{figure:Plots}~e). Here, the dashed red line shows the axial motion in the absence of the cyclotron excitation, and the solid line shows the actual axial motion. The phase shift is $\Delta\phi=\Delta\omega_zt_{ex}$. The particle continues to freely evolve until a predetermined time when the phase shift is read out. There are several steps to this process, designed to minimize the contribution of the detector's Johnson noise to the phase uncertainty without introducing additional phase noise. First, the axial motion and magnetron motions are parametrically amplified using a rf pulse at $\omega_\textrm{rf}=\omega_z-\omega_-$ during the period labeled ``Amplify''.  The amplitude in the axial mode is then rapidly cooled by bringing the resonant circuit into resonance with the axial mode in the cooling configuration, during the period marked ``Cool''. The magnetron mode remains unaffected during this period. The magnetron and axial amplitudes are now exchanged using a $\pi$-pulse at a frequency $\omega_\textrm{rf}=\omega_z+\omega_-$, during the time marked ``$\pi$''. The time between the start of the axial cooling and the start of the $\pi$-pulse is carefully tuned so that unwanted frequency shifts from the amplification of the axial mode in the presence of the magnetic bottle are canceled by the magnetron evolution, discussed further in Section \ref{sec:fduringAmp}. This maneuver results in a final amplified axial amplitude which retains the phase information of the axial excitation before amplification, but has a significantly larger magnitude. Now, during the stage marked ``Detect'', switch 1 between the amplifier and the detection circuit is closed, and the axial phase and amplitude are read out in a phase-sensitive measurement \cite{Cornell1990ModeCrossing}. Finally, switch 1 is opened again to disconnect the amplifier from the LCR circuit, and both axial and magnetron modes are cooled using sideband drives to the ground state in preparation for the next cycle.  

An indicative histogram from repeated measurements of the axial phase can be seen in {Fig.~\ref{figure:Plots}~f)}. In both cases, the mean axial phase when no photons are incident onto the cavity has been subtracted.  Here, the yellow histogram corresponds to the situation when no microwave photon is absorbed by the electron. There is a distribution of phases rather than just a constant phase because of the phase noise associated with the axial zero point fluctuations, even after it is cooled to the ground state. The blue histogram is generated from points where a microwave photon is absorbed by the electron. In this case, the final phase is the convolution of two contributions, one Gaussian distributed noise contribution from the axial zero point fluctuations and a second from the phase shift due to the transient occupation of the $n_+=1$ excited state, which approximately follows an exponential distribution because of the distribution in lifetimes. We can see that in this hypothetical case, if we set a cutoff phase at around 0.75 degrees (dashed line in Fig.~\ref{figure:Plots}~f)), we can conclude that phases higher than this must correspond to a photon absorption event, with a certain false positive rate, while phases less than 0.75 degrees mostly correspond to no photon absorption, with a certain false negative rate. Changing the location of the cutoff reduces the false positive rate, which is a form of readout noise or dark counts, at the expense of detection efficiency. The distributions shown in this figure are purely for illustration, and in the course of the paper, we will derive their true form, so that we can accurately characterize the overall detection probability. In the remainder of the paper, we analyze this detection method in detail to calculate the achievable detection probability. Although we typically analyze the photon counting method in a relatively abstract fashion, without reference to a particular trap geometry or choice of experimental parameters, it is helpful to evaluate the expressions using a particular choice of realizable parameters to make sure the method remains feasible. These reasonable experimental values are listed in Appendix \ref{Appendix:trapParams} and are used throughout unless otherwise stated. The reason governing the choice of parameters will be elucidated in the paper. We now proceed to analyze the photon detection sequence in detail.

\section{Cavity-cyclotron mode interaction}
\label{sec:4}
We now consider the interaction between the trapped electron and the field of the cylindrical cavity formed by the Penning trap electrodes. We note that recently, a closely related system consisting of an electron inside a cylindrical waveguide has also been solved \cite{zdsy-sxcl}. The quantum mechanical evolution of the cyclotron and cavity states, including the effect of losses from the cavity, can be treated using the Lindblad master equation for the density matrix $\varrho$
\begin{align}
    \dot{\varrho}=-\frac{i}{\hbar}[H,\varrho]+\mathcal{L}_D(\varrho )\, ,
    \label{eqLinblad}
\end{align}
where $\mathcal{L}_D(\varrho )$ is the dissipative part defined below, and $H$ is the Hamiltonian of the system. The cyclotron mode is assumed to only decay via interactions with the cavity electric field, other decays are neglected. We determine the form of the interaction Hamiltonian before solving Eq.~\ref{eqLinblad} to characterize the probability of absorption and the lifetime of an electron in the cyclotron excited state. We then consider how short-lived occupations of the excited cyclotron state can be measured by means of phase-sensitive detection of the axial mode, and examine how phase shifts caused by photon absorption can be resolved. Finally, we return to consider the total photon absorption probability of the proposed Penning trap photon counter.  
\subsection{Form of the Hamiltonian}

Ignoring the axial and magnetron modes, the Hamiltonian $H=H_2+H_\textrm{int}$ is the sum of the Hamiltonian $H_2$ associated with the cavity and the cyclotron mode,
\begin{align}
    H_2=\hbar\omega_+(a_+^\dagger a_++\tfrac{1}{2})+\hbar\omega( a_\gamma^\dagger a_\gamma+\tfrac{1}{2})\,,
\end{align}
and an interaction Hamiltonian $H_\textrm{int}$ between the electromagnetic vector potential $\boldsymbol{A}$ of the cavity mode and the electron's momentum $\boldsymbol{p}$ 
\begin{eqnarray}
    H_\textrm{int}=\frac{e}{m}\boldsymbol{A}\cdot \boldsymbol{p} \, .
    \label{equationHint1}
\end{eqnarray}
Assuming that the cavity is designed so that only a single cavity mode interacts with the electron, we can expand $\boldsymbol{A}$ in terms of the field creation and annihilation operators $a_\gamma^\dagger$, $a_\gamma$ for that particular mode, so that in the Heisenberg picture the operator becomes  
\begin{align}
    \boldsymbol{A}(t)&= \sqrt{\frac{\hbar }{2\varepsilon _0\omega \tilde{V}}} \left(\boldsymbol{\epsilon}_\textrm{p}(\boldsymbol{r})a_\gamma e^{-i\omega t} + \boldsymbol{\epsilon}_\textrm{p}(\boldsymbol{r})^*a_\gamma^{\dagger } e^{i\omega t}\right)\, .
\end{align}
Here $\boldsymbol{\epsilon}_\textrm{p}(\boldsymbol{r})$ is the normalized vector potential of the mode in question at the electron position, taken as the origin, and $a_\gamma=a_\gamma(0)$, $a_\gamma^{\dagger}=a_\gamma^\dagger(0)$ are the values of the operators at $t=0$. For the interaction between the cyclotron mode close to the ground state, where the cyclotron radius is small compared to the wavelength, and the TE$_{11q}$ cavity field, where $q$ is odd and the mode is centered at the electron position, we can ignore the spatial dependence and write $\boldsymbol{\epsilon}_\textrm{p}(\boldsymbol{r})=\boldsymbol{\epsilon}_\textrm{p}$, where $\boldsymbol{\epsilon}_\textrm{p}$ is a linear polarisation in the $xy$ plane. For more complex modes, $\boldsymbol{\epsilon}_\textrm{p}(\boldsymbol{r})$ can be evaluated using the standard expressions for the fields in a cavity, see Ref. \cite{Jackson1998ClassicalElectodynamics}. The field is normalized so that the total energy in the cavity volume $V$ is equal to that of a single photon $\hbar\omega=\frac{1}{2}\int d V \, \left(\varepsilon _0 \textbf{E}\cdot\textbf{E}+\frac{1}{\mu _0}\textbf{B}\cdot\textbf{B} \right)$, and the effective mode volume is given by 
\begin{align}
\tilde{V}=\frac{\int \left|E|^2\textrm{d}V\right.}{\left|E_p|^2\right.},    
\end{align}
 where $E_p$ is measured at the position of the trapped electron. 

The cyclotron mode is purely radial, so we can write  $\boldsymbol{p}=m\boldsymbol{\dot{\rho}}$ where $\boldsymbol{\dot{\rho}}$ is the radial velocity in the $x$-$y$ plane. This can be written \cite{Brown1986GeoniumTrap} in the form 
\begin{align*}
    \boldsymbol{\dot{\rho }} =\frac{\boldsymbol{V}^+ \omega _+-\boldsymbol{V}^- \omega _-}{\omega _+-\omega _-}\, .
\end{align*}
The time dependence of the two vectors, $\boldsymbol{V}^\pm$ is $e^{i\omega_\pm t}$. As the cavity operates at around $\omega_+$ we can neglect the $\boldsymbol{V}^-$ term to write 
\begin{align}
    H_{\text{int}}=\sqrt{\frac{e^2\hbar }{2\varepsilon _0\omega \tilde{V}}}\frac{\omega_+\left(a_\gamma  e^{-i\omega_+ t} + a_\gamma^{\dagger } e^{i\omega_+ t}\right)}{\omega _+-\omega_-}\boldsymbol{\epsilon}_\textrm{p}\cdot\boldsymbol{V}^+ \,. 
    \label{equationHint}
\end{align}
The velocity vector $\boldsymbol{V}^+$ can itself be written \cite{Brown1986GeoniumTrap} in terms of the raising and lowering operators $a_+^\dagger$ and $a_+$ of the modified cyclotron mode.  To illustrate a simple case, we consider a linearly polarized field mode along the $x$-axis, and use the fact that $\omega_+\gg\omega_-$ and $\omega_+\simeq\omega$. Then, neglecting the far-off-resonant terms, this becomes
\begin{align}
    H_{\text{int}}=\frac{e\hbar
}{\sqrt{4m \varepsilon _0\tilde{V}}}(a_+^{\dagger }a_\gamma+a_+a_\gamma^{\dagger })\, .
\end{align}

The matrix elements can be written using a basis $\left.\left|n_+\right.,n_{\gamma }\right\rangle$ where $n_{\gamma }$ represents the number of photons in the cavity and $n_+$ the cyclotron state of the electron    
\begin{align}
\left\langle n_++1,n_{\gamma }-1\left|H_{\text{int}}\right|n_+,n_{\gamma }\right\rangle =e\hbar\sqrt{\frac{n_{\gamma }(n_++1)}{4 m\varepsilon _0\tilde{V}}}\, ,
\label{eqnMatrixElement1}
\end{align}
and
\begin{align}
\left\langle n_+-1,n_{\gamma }+1\left|H_{\text{int}}\right|n_+,n_{\gamma }\right\rangle =e\hbar \sqrt{\frac{n_+(n_{\gamma }+1)}{4 m\varepsilon _0\tilde{V}}}\, .
\label{eqnMatrixElement2}
\end{align}
In Appendix A, we confirm that defining the matrix elements in this way leads to the familiar formula for the emission rate of an electron in free space. In the specific case where there is at most one photon in the cavity and the modified cyclotron mode is either in the ground or first excited state, the matrix elements become 
\begin{align*}
\left\langle 1,0\left|H_{\text{int}}\right|0,1\right\rangle =\left\langle 0,1\left|H_{\text{int}}\right|1,0\right\rangle = e\hbar\sqrt{\frac{1}{4 m\varepsilon _0\tilde{V}}}\, ,
\end{align*}
and we can define the coupling rate as 
\begin{align*}
   g= e\sqrt{\frac{1}{4 m\varepsilon _0\tilde{V}}} \, .
\end{align*}
We can now solve the Linblad equation.

\subsection{Solution of the Linblad equation}

\label{sec:linblad}

We consider a restricted basis set $|0\rangle=|0,0\rangle$, $|\gamma\rangle=|0,1\rangle$ \& $|+\rangle=|1,0\rangle$ to investigate either the emission of a single microwave photon into the cavity, or the absorption and subsequent emission of a single photon by a particle initially in the cyclotron ground state. We form the density matrix in the usual way, such that its matrix elements are $\varrho_{ij}=\langle i| \varrho | j \rangle$. The dissipative part of the Linblad superoperator in this case is 
\begin{align}
\mathcal{L}_D(\varrho )=\kappa\left( a_\gamma \varrho a_\gamma^\dagger -\frac{1}{2}\left(a_\gamma^\dagger a_\gamma \varrho + \varrho a_\gamma^\dagger a_\gamma\right)\right),    
\end{align}
where $\kappa=\omega/Q$. This is one of the most basic examples of the interaction between a harmonic oscillator and a quantized field, and it has been studied extensively. For completeness, we provide the solutions in this simple case. The equation can be solved analytically for this restricted basis set. Consider first the situation where the electron is initially $n_+=1$ and all other density matrix elements are zero, and the cavity frequency matches the modified cyclotron frequency $\omega=\omega_+$. The evolution of the cyclotron state is then given by
\begin{align}
    \varrho_{++}(t)=&\frac{e^{-\frac{1}{2}\kappa t} \left(\kappa ^2-8 g^2\right) \cosh \left(\frac{Wt}{2}\right)}{W^2} \nonumber\\&-\frac{e^{-\frac{1}{2}\kappa t} \left(8 g^2-\kappa  W \sinh \left(\frac{Wt }{2}\right)\right)}{W^2} ,
\end{align}
where  $W=\sqrt{\kappa ^2-16 g^2}$. In the weak field coupling limit when $g\ll\kappa$, this gives the familiar form for the Purcell enhancement of the spontaneous emission rate inside a cavity
\begin{align}
    \varrho_{++}(t)&\simeq e^{-\frac{4 g^2 t}{\kappa}}.
    \label{eq:decay}
\end{align}
Interestingly, this quantum calculation of the rate of cyclotron emission exactly matches the classical rate that is calculated from a consideration of the image charges induced on the trap electrodes in the weak coupling limit. This equivalence is demonstrated in Appendix B. Comparing this expression to the free space formula (see Appendix A), we find that in general the spontaneous emission rate can be significantly enhanced by the Purcell factor
\begin{align}
    \frac{4 g^2}{\kappa}  \left(\frac{1}{4\pi\varepsilon _0}\frac{4e^2\omega ^2}{ 3m c^3}\right)^{-1}&=\frac{3 \pi  c^3 Q}{4 V \omega ^3}\gg 1 . \label{eq:decayRate}
\end{align}

Inserting some typical numbers from Appendix \ref{Appendix:trapParams} into the expression for the decay rate $\gamma_q=\frac{4 g^2}{\kappa}$ valid when $g\ll\kappa$ we find $\tau=1/\gamma_q=1.6$ \unit{\milli \second} for $\omega=2\pi\times 30$~\unit{\giga \hertz} for a single electron in a cylindrical cavity of radius 3.7 mm and length 24.55 mm in the $\textrm{TE}_{113}$ mode with $\tilde{V}=2.4\times10^{-7}$ \unit{\meter \cubed} and cavity quality factor $Q=20,000$. This lifetime is very short compared to the timescales over which an image current measurement of the electron motion is typically performed. For instance, in Ref. \cite{DUrso2005Single-particleOscillator}, the authors describe a self-excited oscillator method to detect small frequency shifts for a trapped electron. Depending on the precise method used, the bandwidth varies from 150 \unit{\hertz} to 8 \unit{\hertz}, implying response times above 6 \unit{\milli\second}. It is possible that significantly faster measurements can be performed \cite{Fan2025HighlyDetection}. Our approach is to use a method that does not require the electron to be measured in the excited state, and will be described fully in the next section.

\begin{figure}[htb]
\centering
\includegraphics[]{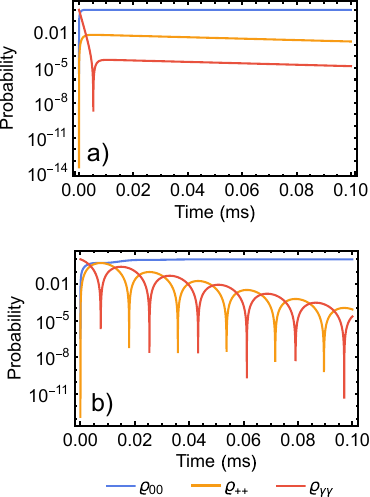}
\caption{Evaluation of the populations in the absolute ground state, excited cavity state and excited cyclotron state given by Eqs. \ref{eq:populations1}-\ref{eq:populations}. Parameters are a) $\omega = 2 \pi \times 30$ \unit{\giga \hertz}, $\tilde{V}=1.2 \times 10^{-7}$ \unit{\meter\cubed}, $Q=10^5$;  b) $\omega = 2 \pi \times 30$ \unit{\giga \hertz}, $\tilde{V}=2.4 \times 10^{-8}$~\unit{\meter\cubed}, $Q=10^6$.}
\label{figure:DensityMatrixPlots}
\end{figure}

We now consider the situation in which the system begins in the $|0,1\rangle$ state with one photon in the cavity, which is relevant for photon counting. The solution of the Master equation for the diagonal elements of the density matrix corresponding to the probability for the system to be found respectively in the ground state, $n_+=1$ and $n_\gamma=1$ states are:
\begin{align}
    \varrho_{00}(t)=& 1-\varrho_{++}(t)-\varrho_{\gamma\gamma}(t) \label{eq:populations1}\\
    \varrho_{++}(t)=&\frac{16 g^2 e^{-\frac{1}{2} \kappa  t} \sinh ^2\left(\frac{W t}{4}\right)}{W^2} \label{eq:populations2}\\
    \varrho_{\gamma\gamma}(t)=&\frac{e^{-\frac{1}{2} \kappa  t} \left(\kappa ^2-8 g^2\right) \cosh \left(\frac{W t}{2}\right)}{W^2}\nonumber\\
    &-\frac{e^{-\frac{1}{2} \kappa  t} \left(8 g^2+\kappa W \sinh \left(\frac{W t}{2}\right)\right)}{W^2}
    \label{eq:populations}
\end{align}
The resulting probabilities are plotted in Fig.~\ref{figure:DensityMatrixPlots}, and the parameters used in these evaluations are listed in the caption. Fig.~\ref{figure:DensityMatrixPlots}~a) shows the weak coupling limit with $g\ll \kappa$. In this case, the initial excitation in the cavity, $\varrho_{\gamma\gamma}$, mostly decays quickly away, while a small fraction, around 1\%, is excited to $n_+=1$, which decays more slowly. Fig.~\ref{figure:DensityMatrixPlots}~b) shows the onset of strong coupling with $\kappa \simeq g$. In this case, nearly 50\% of the population is transferred into $n_+=1$, and then it subsequently oscillates between $n_+=1$ and $n_\gamma=1$ while slowly decaying. Increasing $g$ further with respect to $\kappa$ further increases the height of the initial peak in $\varrho_{++}$ so that it approaches 100\%. 
The first amplitude peak, where the probability for the photon to transfer to the excited state is maximized, is found from the first maximum of Eq.~\ref{eq:populations2}, and depends only on the ratio $S=\frac{\kappa}{4g}$ 
\begin{align}
    P_e&=(S+\sqrt{S^2-1})^{-\frac{2S}{\sqrt{S^2-1}}} ,
    \label{eq:maximum}
\end{align}
which occurs at a time
\begin{align}
    t&=\frac{4}{W} \ln \left[\frac{\kappa+W}{2  g}\right]\, .
\end{align}

Equation \ref{eq:maximum} suggests that the cavity will need to have as high a quality factor as possible, ideally as close to the maximum set by the frequency width of the axion signal, $Q=Q_a=10^{6}$, to make $g$ as large as possible with respect to $\kappa$. This is much larger than the $Q$-factors achieved in Penning traps that double as microwave cavities, which  typically have around  $Q=5,000$ 
for the TE modes. Increasing the $Q$-factor beyond these values is the major technical challenge of our method.  At these frequencies and cryogenic temperatures, cavities made of normal metals such as copper are limited to around $Q=4\times 10^4$ \cite{9699394}. 
Cavities coated in superconducting NbSn$_3$ have reached $Q=1\times 10^5$ at 1 T, falling to $Q=6\times 10^4$ at 2 T \cite{9699394}. Superior performance can be achieved with a hybrid copper/ NbTi cavity coating, achieving $Q=6\times 10^5$ at 1 T, falling to $Q=4\times 10^5$ at 2 T for an unloaded cavity \cite{Alesini2019GalacticCavity}. Both these cavities were measured at resonant frequencies close to 9 GHz, so we would expect the $Q$-factor to be lower at 30-60 GHz. Even higher quality factors can be reached using thin superconducting NbTi coatings on a thick, low-loss substrate like alumina \cite{Xi2010Far-InfraredMa} on all surfaces parallel to the magnetic field lines. By these methods, we think $Q$-factors in the range $Q=10^5-10^6$ could eventually be achieved. In this paper we consider $Q=10^4-10^6$. 

\begin{figure}[htb!]
\centering
\includegraphics[width = 0.45\textwidth]{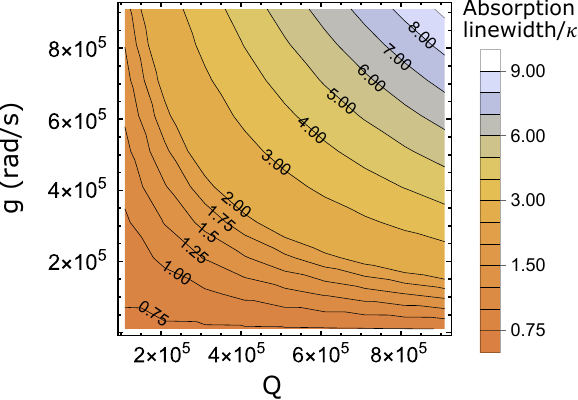}
\caption{The absorption linewidth of the electron and cavity when a single electron is placed in a cavity, divided by the cavity linewidth $\kappa$. Here $\omega=2\pi\times30$~GHz.}
\label{figure:Detuning}
\end{figure}

So far, we have assumed that the cavity frequency exactly matches the modified cyclotron frequency. In the sections that follow, we will see how the motion of the electron can introduce frequency shifts which may lead to a relative detuning $\Delta=\omega-\omega_+$ between these frequencies. To understand how this affects the electron-cavity coupling, we numerically calculate $P_e$ for a range of detunings $\Delta$ and then find the full width at half maximum of  $P_e$ with respect to $\Delta$, which we call the absorption linewidth of the electron-cavity system. The electron-cavity linewidths obtained are shown in Fig.~\ref{figure:Detuning}. In this figure, we have divided the electron-cavity linewidth by $\kappa$, the cavity linewidth without an electron. The contours represent the extent to which the electron-cavity linewidth is broader or narrower than the cavity's linewidth  $\kappa$. For the electron photon counter, the typical operating parameters are in the bottom section of this figure, in the low $g$ region, where we can approximate the absorption linewidth by the cavity linewidth $\kappa$.

We now consider the phase accrued by the electron as a result of occupying the excited state. First, to get a sense of the typical magnitude of the effect, we calculate the mean phase shift, $\langle\phi\rangle=\Delta\omega_z\langle t_{ex}\rangle$. This can be calculated from the average time spent in the cyclotron excited state $\langle\phi\rangle=\Delta\omega_z\int_0^\infty\varrho_{++}(t)dt=\frac{\Delta\omega_z}{\kappa}$. Interestingly, this mean shift is independent of $g$. For the parameters in Table \ref{tab:placeholder}, with $Q=10^6$, $\langle\phi\rangle=1$ degree, indicating the typical size of phase shifts that will need to be resolved.  

To proceed further, we need to derive the form of the probability density function (PDF) $p_{\phi}(\phi)$ for the phase accrued. Once a decay into the $|0\rangle$ state occurs, the excitation is gone from the cavity-electron system and can no longer contribute to the final axial phase via the continuous Stern-Gerlach effect. Given that the system definitely starts with an excitation in the cavity, and it can only decay into the $|0\rangle$ state from the cavity excited state, the PDF for the time $t$ to decay into the $|0\rangle$ state is 
\begin{align}
p_t(t)=\kappa\varrho_{\gamma\gamma}(t)   \,. \label{eq:ProbvsT}
\end{align}
The extra phase accumulated by the system in the time $t$ while it remains in the excited state is 
\begin{align}
    \phi=\Phi(t)=\Delta\omega_z \int_0^t\frac{\varrho_{++}(t')}{\varrho_{++}(t')+\varrho_{\gamma\gamma}(t')}dt' \, . \label{eq:PhivsT} 
\end{align}
We can use Eq.~\ref{eq:PhivsT} to transform $p_t(t)$ into $p_\phi(\phi)$ using the normal rules for a change of variables of a PDF
\begin{align}
    p_{\phi}(\phi)&=p_t(\Phi^{-1}(\phi))\left|\frac{\textrm{d}t}{\textrm{d}\phi}\right|\nonumber\\&=p_t(\Phi^{-1}(\phi))\frac{\varrho_{++}(\Phi^{-1}(\phi))+\varrho_{\gamma\gamma}(\Phi^{-1}(\phi))}{\Delta\omega_z\varrho_{++}(\Phi^{-1}(\phi))},
\end{align}
where $\Phi^{-1}(\phi)$ is the inverse function of $\Phi(t)$. The probability distribution $p_\phi(\phi)$ should be folded into an interval spanning $2\pi$. However, restricting to a single photon in the cavity, a realistic electron single-photon counter rarely results in a phase exceeding $\pi$, with a single absorption. Where multiple photons are present, the effects of phase wrapping must be considered.

\begin{figure}[htb]
\centering
\includegraphics[]{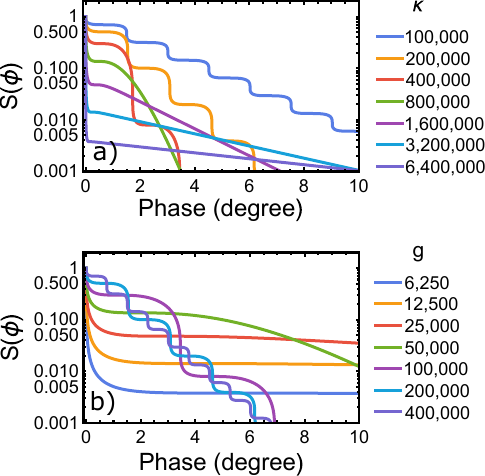}
\caption{a) Survival function $S(\phi)$ for $g=2\times10^5$ and various values for $\kappa$. b) Survival function $S(\phi)$ for $\kappa=2\times10^5$ and various values of $g$. In both graphs $\omega = 2 \pi \times 30$ \unit{\giga \hertz} and $B_2=10^5$, other parameters as in Appendix \ref{Appendix:trapParams}. The green curves in each graph corresponds to the boundary between under-damped cyclotron-cavity population oscillations and overdamped decay.}
\label{figure:PDFCDF}
\end{figure}

More informative than the PDF is the survival function $S(\phi)=1-C(\phi)$ where $C(\phi)$ is the cumulative density function (CDF) $C(\phi)=\int_0^{\phi}p_\phi (\phi')d\phi'=\Delta\omega_z\int_0^{\Phi^{-1}(\phi)}p_t (t')dt'$. The survival function $S(\phi)$ gives the fraction of the distribution which has a phase greater than $\phi$. There will typically be a phase uncertainty $\sigma_\phi$ associated with measuring the axial phase, which implies that $\phi$ should exceed some minimum value $\phi_m$ in order for the phase shift to be unambiguously detected. Hence, we are interested in maximizing $S(\phi_m)$. Figs.~\ref{figure:PDFCDF}~a) and b) plot $S(\phi)$ for various values of $g$ or $\kappa$ respectively. Considering first \ref{figure:PDFCDF} a), we find that generally, for small phases, $S(\phi)$ is maximized if $\kappa$ is made small as possible. At larger phases, it can sometimes be beneficial to choose a larger value of $\kappa$, e.g. at $\phi=4$ degrees $\kappa=1.6\times10^6$ outperforms $\kappa=4\times10^5$. Fig.~\ref{figure:PDFCDF}~b) shows that if $\kappa$ is fixed, the optimum value of $g$ depends on the choice of $\phi_m$. If the phase threshold for detection is small, then generally it is beneficial to have $g\gg \kappa$. However, if $\phi_m$ is large, it may be better to have $g$ of similar size to or smaller than $\kappa$. As a general rule, for high efficiency detection, $\kappa$ should be as small as possible and $g$ should be optimized once the phase resolution associated with the detection of the axial phase is known.          


In this section, we have derived a method for calculating the phase shift from a transient occupation of the cyclotron level, and how this depends on $g$ and $\kappa$. To calculate if this phase is resolvable, we need to know the phase noise associated with a measurement of the axial mode. In the next section, we consider the dynamics of the axial and magnetron motion, and use these results to derive the limits of axial phase resolution and hence the overall detection probability. 

\section{Axial and magnetron motion}
\label{sec:5}
\subsection{Wigner function description of the motion}
 As we are interested in tracking the motion of an electron that begins in the absolute ground state, we must initially treat the motion quantum mechanically. The interaction Hamiltonians for the excitation, amplification, and coupling pulses are found by using $H_{int}=e\nabla\hat\varphi$ with the electrostatic potential $\hat\varphi$ operator chosen appropriately for electric fields $\boldsymbol{E}_d = \mathcal{E}_d \cos(\omega_z t) \hat{\boldsymbol{z}}$ for the excitation pulse and  $\boldsymbol{E}_{i} = \mathcal{E}_i \cos(\omega_{i}t) ( x \hat{\boldsymbol{z}}+ z \hat{\boldsymbol{x}})$, where $i=p$,  $\omega_{p}=\eta+\omega_z-\omega_-$ during the amplification, $i=\pi$, $\omega_{\pi}=\eta+\omega_z+\omega_-$ during the coupling pulse, and $\eta$ represents a possible frequency detuning. Where operators $\hat x$ and $\hat z$ appear, these are replaced by functions of the creation and annihilation operators for the axial and magnetron mode using the expressions in Ref. \cite{Brown1986GeoniumTrap}. The Hamiltonians are, for the excitation pulse 
\begin{align}
\hat H_d&=\frac{e|\mathcal{E}_\textrm{d}|}{2} \sqrt{\frac{\hbar }{2 m \omega _z}} \left(a_z^{\dagger }e^{-i \text{$\omega_z $t}}+a_ze^{i \text{$\omega_z $t}}\right) ,
\end{align}
for the amplification pulse
\begin{align}
    \hat H_p=\frac{ie \hbar |\mathcal{E}_p|}{4 m \sqrt{\left(\omega _+-\omega _-\right) \omega _z}}  \left(a_-  a_ze^{i \text{$\omega_p $t}}- a_-^\dagger a_z^\dagger e^{-i \text{$\omega_p $t}} \right),
\end{align}
 and for the coupling pulse
\begin{align}
    \hat H_\pi=\frac{ie \hbar |\mathcal{E}_\pi|}{4 m \sqrt{\left(\omega _+-\omega _-\right) \omega _z}} \left(a_-  a_z^\dagger e^{i \text{$\omega_\pi $t}}-a_-^\dagger a_z e^{-i \text{$\omega_\pi $t}} \right).
\end{align}

We now describe the evolution of the system under the sequence of pulses described in Section \ref{sec:overview}. The change of the state under the excitation pulse can be described using standard unitary operators, however, the amplification and remaining steps in the sequence are best treated using Wigner functions. During our analysis of these steps, we introduce various standard results from Refs. \cite{Mollow1967QuantumI,Mollow1967QuantumII} for Wigner functions as required.  

\subsubsection{Excitation}
The excitation pulse Hamiltonian causes the state to evolve under the displacement operator $\mathcal{D}(\alpha)=e^{-i \hat H_d t_d/\hbar}$, where $\alpha=\frac{-ie\mathcal{E}_dt_d}{2} \sqrt{\frac{1}{2 \hbar m \omega _z}}$. We label the state of the axial and magnetron modes by their complex coherent state amplitudes $\alpha$ and $\beta$ for the axial and magnetron mode, respectively $|\psi\rangle=|\alpha\beta\rangle$, where $a_z|\alpha\beta\rangle=\alpha|\alpha \beta\rangle$ and $a_-|\alpha\beta\rangle=\beta|\alpha \beta\rangle$. We label the coherent axial amplitude state created by the displacement $\mathcal{D}(\alpha_i)|00\rangle=|\alpha_i 0\rangle$ when acting on the particle which is initially in the ground magnetron and axial state. 

\begin{figure*}[t]
    \centering
    \includegraphics[]{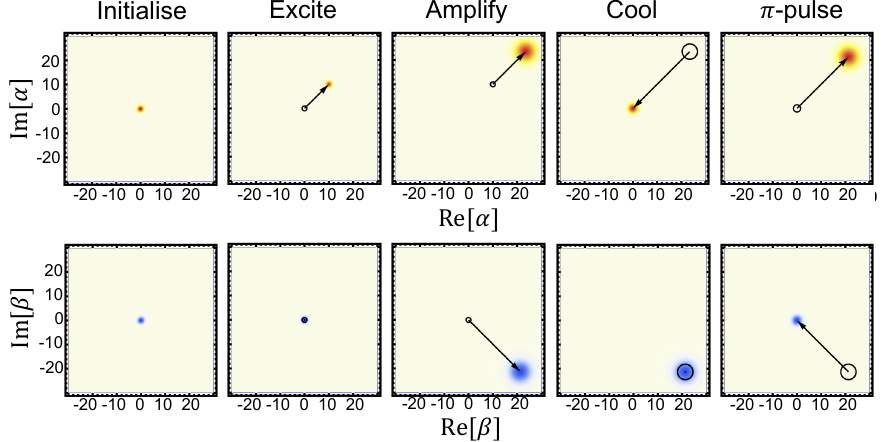}
    \caption{Wigner function during the photon counting sequence for the reduced density matrix for the axial and magnetron modes, expressed in terms of the real and imaginary parts of the coherent amplitudes $\alpha$ and $\beta$, respectively that appear in the definition of the Wigner function, Eq.~\ref{eq:twoModeAmp}. The black circle shows the standard deviation of the Wigner function at the preceding step for comparison.
    \label{fig:wignerGraph}}
\end{figure*}

\subsubsection{Free evolution}
During the free evolution following the axial excitation, the state gains a phase factor $\phi_T=\omega_z t_\textrm{ev}+\Delta\omega_zt_\textrm{ex}$, where $\Delta\omega_z=0$ if the cyclotron state remains unchanged. The state just before the amplification pulse is therefore $|\alpha' 0\rangle$ where $\alpha'=e^{-i\phi_T}\alpha_i$.   
\subsubsection{Amplification}
The dynamics of a system which begins in a thermal or coherent state and undergoes evolution according to the two-mode non-degenerate parametric amplification Hamiltonian $\hat{H}_a$ has been studied extensively by Mollow and Glauber \cite{Mollow1967QuantumI,Mollow1967QuantumII}, and we reproduce much of the following analysis from their work. We define $k=\frac{e \hbar |\mathcal{E}_a|}{4 m \sqrt{\left(\omega _+-\omega _-\right) \omega _z}}$ and adjust the initial time to remove the complex pre-factor. We note that as the magnetron energy is negative, to match our expression with Refs. \cite{Mollow1967QuantumI, Mollow1967QuantumII} we would need to set $\omega_a=\omega_z$ and  $\omega_b=-\omega_-$, explaining why in our case parametric amplification occurs at $\omega_z-\omega_-$ rather than  $\omega_z+\omega_-$. The time evolution is solved in the Heisenberg picture, where the time-dependent creation operators can be expressed in terms of the time-independent creation operators and various hyperbolic functions according to 
\begin{align}
    a_z(t)&=a_zc_z(t)+a_-^\dagger s_z(t),\\
    a_-(t)&=a_-c_-(t)+a_z^\dagger s_-(t),
\end{align}
where
\begin{align}
    c_z(t)&=e^{-i\omega_zt}\cosh(kt),\\
    s_z(t)&=ie^{-i\omega_zt}\sinh(kt),\\
    c_-(t)&=e^{i\omega_-t}\cosh(kt),\\
    s_-(t)&=ie^{i\omega_-t}\sinh(kt).
\end{align}
Since we want to make a connection between the quantum description of the system and a classical phase space probability distribution, we use the Wigner function to keep track of the state of the system. The Wigner function for this two-mode system is given by 
\begin{align}
W(\alpha, \beta, t) &= \frac{1}{\pi^4} \int e^{\alpha \eta^* + \beta \xi^* - \text{c.c.}} \chi (\eta, \xi, t) \, d^2\eta \, d^2\xi,
\label{eq:twoModeAmp}
\end{align}
where the characteristic function is
\begin{align}
    \chi(\eta, \zeta, t) &= \text{Tr} \left\{ \rho(t) e^{\eta a_z^{\dagger} + \zeta a_-^{\dagger} - \eta^* a_z - \zeta^* a_-} \right\}.
\end{align}
For our system, the Wigner function after amplification for a time $t_a$ is given in terms of the functions $\alpha_{0,c}$, $\beta_{0,c}$
\begin{align}
W(\alpha, \beta, t)&=W(\alpha_{0,c}(\alpha,\beta,t_a),\beta_{0,c}(\alpha,\beta,t_a),0)\nonumber\\&=
\left(\frac{2}{\pi }\right)^2e^{-2\left|\alpha_{0,c}(t_a) -\alpha'\right|^2}e^{-2\left|\beta_{0,c}(t_a) \right|^2},
\end{align}
where $\alpha_{0,c}(t)=\alpha c_z^*(t)-\beta ^* s_{-}(t)$ and $\beta_{0,c}(t)=\beta c_-^*(t)-\alpha^* s_{z }(t)$. The Wigner function for the reduced density matrix is plotted in Fig.~\ref{fig:wignerGraph}. Here, the top row corresponds to the axial mode (coherent amplitude $\alpha$) and the bottom row to the magnetron mode (coherent amplitude $\beta$). The two-mode Wigner function begins with both modes in their ground state. After the excitation drive, it can be written as the product of two exponentials, which are functions of only $\alpha$ and $\beta$ respectively, corresponding to the product of a coherent axial state of amplitude $\alpha'$ and the magnetron ground state. The parametric drive couples together the axial and magnetron modes, as shown in the ``Amplify'' step in Fig.~\ref{fig:wignerGraph}, and prevents the joint Wigner function from being expressed as a separable product of functions which depend on only $\alpha$ and $\beta$ respectively. The amplitudes become increasingly correlated as the parametric drive continues. This follows from the fact that the Heisenberg operator $\hat{X}(t)= a_z^\dagger(t) a_z(t)-a_-^\dagger(t) a_-(t)$, representing the amplitude difference, is a constant of motion \cite{Mollow1967QuantumI}. Its expectation value can be calculated at $t=0$,
\begin{align}
  \langle a_z^\dagger a_z-a_-^\dagger a_-\rangle&=\int\int (|\alpha|^2-|\beta|^2 )W(\alpha,\beta,t)d^2\alpha d^2\beta\nonumber\\&=|\alpha'|^2.
\end{align}
As the individual amplitudes grow coherently according to hyperbolic functions, but their amplitude difference remains constant, the fractional amplitude difference becomes closely correlated.

Functions of $\hat{X}$ are also constants of motion. This means the standard deviation of the amplitude difference $(\langle\hat{X}^\dagger(t)\hat{X}(t)\rangle-\langle\hat{X}(t)\rangle^2)^{1/2}$ does not grow, and can be calculated at the start of the amplification pulse as  
\begin{align}
    \sqrt{\langle \hat{X}^{\dagger}\hat{X}\rangle -\langle \hat{X}\rangle ^2}=|\alpha'|.
    \label{eq:amplitudeVariance}
\end{align}
This helpful correlation will enable the most troublesome frequency shifts imposed by the magnetic bottle during the detection period to be removed.

We note that, despite the amplification, the magnetron radius is small compared to the axial oscillation amplitude after the parametric amplification pulse is completed. The absolute value of the most probable magnetron radius at this point is
\begin{align}
    |\rho_-|&=\sqrt{2}\rho_0\bar{\beta}= \sqrt{\frac{2 \hbar }{m \omega _z}}\sqrt{\frac{2 \omega _-}{\omega _z}}s^2(t_a)|\alpha'|,
\end{align}
where $\rho_0=\sqrt{\frac{2 \hbar\omega _- }{m \omega _z^2}}$. This can be compared to the most probable axial amplitude after amplification   
\begin{align}
    |z|&=\sqrt{2}z_0\bar{\alpha}= \sqrt{\frac{2 \hbar }{m \omega _z}}c^2(t_a)|\alpha'|,
\end{align}
where $z_0=\sqrt{\frac{\hbar }{m \omega _z}}$. Note that for large amplifications where $c^2(t_a)\simeq s^2(t_a)$, the magnetron radius remains smaller by a factor $\sqrt{\frac{2 \omega _-}{\omega _z}}$, which means that in most geometries, the magnetron mode will be well away from the trap walls.

\subsubsection{Axial cooling}

When the axial resonant circuit is connected in its cooling mode, the effect is to rapidly damp the axial motion, as pictured in Fig.~\ref{fig:wignerGraph}. The resulting Wigner function can be expressed as the product of a thermal state for the axial mode, while the magnetron mode is the Wigner distribution associated with the reduced density matrix $\varrho_B=\textrm{Tr}_A\{\varrho\}$, where the axial state vectors have been traced over. The resulting Wigner function after a time $t_c$ when the axial cooling is complete is 
\begin{align}
    W(\alpha, \beta)&=\frac{2e^{\frac{-\left|\alpha \right|^2}{\langle n_z\rangle +\frac{1}{2}}}}{\pi^2(\langle n_z\rangle +\frac{1}{2})}\int e^{-2\left|\beta -\beta '|^2\right.}\textrm{P}(\beta',\bar{\beta} )d^2\beta'.
    \label{eq:wignerAfterCooling}
\end{align}
Here, $\langle n_z\rangle$ is the average axial quantum number after the mode is cooled to $T_{e}$. We have left the portion of the Wigner function for the magnetron mode in terms of the $\textrm{P}$-function, which can be used to expand the reduced density matrix in terms of the coherent states of the magnetron mode $|\beta\rangle$ according to
\begin{align}
    \varrho_B &= \int \textrm{P}(\beta,\bar{\beta} ) |\beta\rangle \langle \beta| \, d^2\beta .
    \label{q:varro}
\end{align}
where in our case 
\begin{align}
     \textrm{P}(\beta,\bar{\beta} ) &= \frac{1}{\pi s^2(t_a)}  \exp \left[ -\frac{|\beta - \bar{\beta}|^2}{s^2(t_a)} \right]. 
     \label{eq:pdeff}
\end{align}
Here $s^2(t)=\sinh^2{kt}$, $c^2(t)=\cosh^2{kt}$ and $\bar{\beta} = e^{-i \omega_-t_c}(\alpha')^*s_-(t_a)$. Using the P-representation emphasizes that the density matrix for the system corresponds to the classical phase space probability distribution for complex axial position $\overline{\beta}(t)$. We will spell this connection out more clearly at the end of the detection sequence.

\subsubsection{Axial-Magnetron \mathinhead{\pi}-flip}
We now use a rf pulse at $\omega=\omega_z+\omega_-$ to exchange the axial and magnetron amplitudes. The basic treatment of Refs. \cite{Mollow1967QuantumI,Mollow1967QuantumII} can easily be expanded to deal with the Hamiltonian $H_\pi$. The new solutions for the time-dependent operators under this Hamiltonian are
\begin{align}
    a_z(t)&=a_z \tilde{c}_z(t)+a_-\tilde{s}_z(t),\\ 
    a_-(t)&=a_- \tilde{c}_-(t)+a_z \tilde{s}_-(t),
\end{align}
where
\begin{align}
     \tilde{c}_z&=e^{-i \omega _zt} \cos (k t),\\ 
     \tilde{s}_z&=ie^{-i \omega _zt}\sin (k t),\\
     \tilde{c}_-&=e^{i \omega _-t} \cos (k t),\\ 
     \tilde{s}_-&=ie^{i \omega _-t}\sin (k t).
\end{align}
Once again, the time evolution of the Wigner function can be expressed in terms of the time evolution of two time-dependent functions $\alpha_{0,\pi}$ and $\beta_{0,\pi}$ of the complex variables $\alpha$ and $\beta$
\begin{align}
W(\alpha, \beta, t)=W(\alpha_{0,\pi}(\alpha,\beta,t),\beta_{0,\pi}(\alpha,\beta,t),0),
\end{align}
where in this case the functions are given by $\alpha _{0 \pi}=\alpha\tilde{c}_z^*+\beta\tilde{s}_-^*$ and $\beta _{0,\pi}=\alpha  \tilde{s}_z^*+\beta\tilde{c}_-^* $.

We can see that, up to some phase factors, this corresponds to the periodic exchange of $\alpha\leftrightarrow\beta$ with a Rabi rate $\Omega=2k$. In particular, if we set the pulse duration to give a $\pi$-pulse,  $t=\frac{\pi}{2k}$, then if the initial state is $\alpha=\alpha_0$ and $\beta=\beta_0$ the Wigner function becomes
\begin{align}
W(\alpha, \beta, t=\frac{\pi}{2k})=W(\beta _0 e^{-i \pi  \left(\frac{\omega_-}{2 k}+\frac{1}{2}\right)},\alpha _0 e^{i \pi  \left(\frac{\omega_z}{2 k}-\frac{1}{2}\right)},0).
\end{align}
Using the form of the Wigner function after the axial cooling, Eq.~\ref{eq:wignerAfterCooling}, we have, after the axial-magnetron $\pi$-pulse, 
\begin{align}
    W(\alpha, \beta)&=\frac{2e^{\frac{-\left|\beta\right|^2}{\langle n_z\rangle +\frac{1}{2}}}}{\pi^2(\langle n_z\rangle +\frac{1}{2})}\int e^{-2\left|\alpha -\alpha'|^2\right.}\textrm{P}(\alpha',\bar{\alpha}'')d^2\alpha'.
    \label{eq:wignerAfterPi}
\end{align}
We can write out $\bar{\alpha}''$ fully to show that the final phase depends on $\Delta\omega_z$ as required: $\bar{\alpha}''=\alpha_i^* \exp[-i \phi_f]s_-(t_a)$ with $\phi_f=\pi  (\frac{\omega_z}{2 k}-\frac{1}{2})+\omega_-t_c+\omega_z t_\textrm{ev}+\Delta\omega_zt_\textrm{ex}$

\subsection{Axial detection}

Finally, we now close the switch between the resonant circuit and the axial amplifier and detect the axial amplitude. As the magnetron mode is not detected, we can take a partial trace over the states of the magnetron mode. The resulting single-mode Wigner distribution can be written in the P-representation, using the expression for P from Eq.~\ref{eq:pdeff}
\begin{align}
    W(\alpha)&=\frac{2}{\pi }\int e^{-2\left|\alpha -\alpha'|^2\right.}\textrm{P}(\alpha',\bar{\alpha}'')d^2\alpha '.
    \label{eq:axialDetectionWigner}
\end{align}
This Wigner function has a natural interpretation as a classical probability distribution over coherent states, where each coherent state with amplitude $\alpha'$ is given a weight $\textrm{P}(\alpha',\bar{\alpha}'')$. By using Eq.~\ref{q:varro}, the classical probability density function for the amplitude $|\bar{z}|$ and phase $\phi_z$ of the electron's complex position $\bar{z}=|\bar{z}|e^{i\phi_z}$ corresponding to Eq.~\ref{eq:axialDetectionWigner} is
\begin{widetext}
\begin{align}
    \mathcal{F}(|\bar{z}|,\phi_z)&=\frac{1}{2 \pi  \sigma_{2,z}^2}\exp\left[-\frac{\left(|\bar{z}|\cos(\phi_z)-|z_f| \sin (\phi_f)\right)^2+\left(|\bar{z}|\sin(\phi_z)-|z_f| \cos (\phi_f)\right)^2}{2\sigma_{2,z}^2}\right]\, .
    \label{eq:amped}
\end{align}
\end{widetext}
Here
\begin{align}
    \sigma_{2,z}^2&=\frac{z_0^2}{2}\left( c^2\left(t_a\right)+s^2\left( t_a\right)\right),
\end{align}
and
\begin{align}
    |z_f| &=z_0s(t_a)|\alpha_i|\sqrt{2}.
\end{align}

For the remainder of this paper, we use this classical PDF to analyze the detection noise. This gives a good approximation to all quantities of interest provided that $|z_f|\gg\sigma_{2,z}$. The form of expressions derived from Eq.~\ref{eq:amped} are not correct if the initial excitation amplitude is small, $|\alpha_i|\lesssim\frac{1}{\sqrt{2}}$; however, this is not the situation in the photon counting application. 

The detection system adds Johnson noise, which we incorporate into our analysis classically. When interacting with the detection system, the instantaneous amplitude of the voltage across the impedance $Z(\omega_z)$ of the detection circuit at the frequency $\omega_z$ is

\begin{align}
    \bar{V}_\textrm{f}=\frac{e \omega _z|Z(\omega_z)| |\bar{z}|}{ D_\textrm{eff,z}}\, .
    \label{eq:vtozeta}
\end{align} 
Here $D_\textrm{eff,z}$ is the effective electrode distance for the axial detection, which can be calculated analytically for a typical trap geometry and is of the same order of magnitude as the trap radius. The resulting signals are typically Fourier transformed prior to detection, with a signal bandwidth $2\pi\times\delta\nu=1/\tau$ specified by the measurement time $\tau$. As we will later show, the frequency shifts during detection are typically within the bandwidth $\delta\nu$, which means we can neglect them. During detection, the amplitude decays according to $|\bar{z}(t)|=|\bar{z}_2|e^{-\frac{\gamma_z t}{2}}$, where $\bar{z}_2$ is the axial amplitude at the start of the acquisition time, which means that the average signal amplitude after a time $\tau$ is 

\begin{align}
    \bar{V}_\textrm{f}(\tau)&=\frac{2m \omega _z  D_\textrm{eff,z}|\bar{z}_2|(1-e^{-\frac{\gamma_z\tau}{2}})}{e\tau} \label{eq:correctedEq21}.
\end{align}
The axial decay rate due to the particle-detector interaction is
\begin{align}
    \gamma_z&=\frac{e^2 |Z(\omega_z)|}{m D_{\text{eff},z}^2}.
\end{align}
 The rms value of the voltage amplitude of the detector averaged over at time $\tau$ is given by

\begin{align}
    V_n=\sqrt{\frac{4k_B T_D \textrm{Re}[Z(\omega_z)] }{\tau}+\frac{e_n^2}{\tau}} \, .
\end{align}

Here, the first term in the square root is the Johnson noise from the finite detector temperature $T_D$, and the second term is a possible equivalent input noise $e_n$ intrinsic to the amplifier.

We could, in principle, convert all the axial amplitudes into voltages across the detection circuit and continue our discussion in the voltage/phase space. However, it is more physically meaningful for the discussion of the experimental imperfections in the sections that follow to continue working in terms of particle amplitudes and phases, rather than detector voltages. We therefore use Eq.~\ref{eq:correctedEq21} to convert the detector noise voltage back into an additional fictitious axial amplitude noise. This means we treat the noise sources associated with the detector as an additional, Gaussian distributed vector with random phase and an amplitude chosen such that, when this vector is converted into a voltage using Eq.~\ref{eq:vtozeta}, the resulting voltage noise is $V_n$. The final distribution including particle and detector fluctuations is a displaced Gaussian
\begin{widetext}

\begin{align}
    \mathcal{F}(|\bar{z}|,\phi_z)&=\frac{1}{2 \pi  \sigma_{3,z}^2}\exp\left[-\frac{\left(|\bar{z}|\cos(\phi_z)-|z_f| \sin (\phi_f)\right)^2+\left(|\bar{z}|\sin(\phi_z)-|z_f| \cos (\phi_f)\right)^2}{2\sigma_{3,z}^2}\right]\, 
    \label{eq:finalDist}
\end{align}

\end{widetext}
with 
\begin{align}
    \sigma_{3,z}^2&=\sigma_{2,z}^2+\sigma_D^2\, ,\\
    \sigma _D^2&=\frac{e^2\tau\left( e_n^2+4 k_B \textrm{Re}[Z(\omega_z)] T_D\right)}{4m^2\omega _z^2D_{\text{eff},z}^2(1-e^{-\frac{\gamma_z\tau}{2}})^2}.\label{eq:sigmaD}
\end{align}

This is the final distribution which we work with in the remainder of the paper. Note that, although we derived this expression for a particle cooled to the ground state, essentially the same result follows in the case of a particle with an initial axial temperature $T_z$ and magnetron temperature $T_-$. In this case, we replace $\sigma_{2,z}^2\rightarrow\sigma_\textrm{Th}^2$ where 
\begin{align}
   \sigma_\textrm{Th}^2&=\frac{k_B}{m\omega_z^2}\left( T_z c^2\left(t_a\right)+\frac{\omega_z}{\omega_-}T_-s^2\left( t_a\right) \right).
\end{align}

\subsection{Zero point noise after the initial axial excitation}

\begin{figure*}[t]
\centering
\includegraphics[]{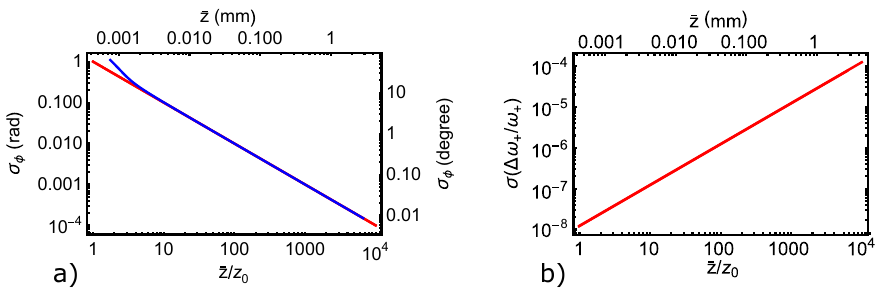}
\caption{a) Standard deviation of measured phases as a function of axial amplitude, blue is calculated numerically from Eq.~\ref{eq:finalDist}, red is the approximation for $\sigma_\phi$ from Appendix \ref{Appendix:standardDeviation}. b) Standard deviation of cyclotron shift as a function of axial amplitude evaluating Eq.~\ref{eq:energyScatter2} using values from Appendix \ref{Appendix:trapParams}.}
\label{figure:PhaseScatterAndAmplitudeNoise}
\end{figure*}

We consider the axial phase scatter immediately before the amplification pulse is applied. In Fig.~\ref{figure:PhaseScatterAndAmplitudeNoise}~a) we plot, in blue, the standard deviation of the axial phase at this time, $\sigma_\phi=\sqrt{\left< \phi^2\right>-\left< \phi\right>\left<\phi\right>}$ as a function of the initial axial excitation amplitude $|z_i|$ calculated numerically from Eq.~\ref{eq:finalDist}. On the lower axis, this is expressed as a ratio of the ground state amplitude $z_0=\sqrt{\frac{\hbar}{m \omega_z}}$ and on the upper axis this is expressed in millimeters using numbers from Appendix  \ref{Appendix:trapParams}. Also plotted in red is the approximation for $\sigma_\phi$ from Appendix \ref{Appendix:standardDeviation}, which is
\begin{align}
  \sigma_\phi&\simeq\sqrt{\frac{\hbar}{m \omega_z\bar{z}^2}}.
  \label{eq:approxScatter1}
\end{align}
This approximation matches the true value when {$\bar{z}/z_0\gg1$}. As expected, increasing the initial axial amplitude results in lower phase uncertainty, so if there were no other considerations, it would be advantageous to use as large a value of $|\bar{z}_i|$ as possible.   

The existence of the $B_2$ term, while crucial for resolving cyclotron phase jumps, also effectively places an upper limit on $|\bar{z}_i|$ and hence the achievable phase resolution. The $B_2$ term causes the cyclotron frequency to depend on the axial amplitude. The frequency shift is $\Delta \omega_+/\omega_+=\frac{\bar{z}^2B_2}{2B_0}$. The direct effect of this shift can be compensated--once the desired excitation amplitude is chosen, the change in cyclotron frequency can be removed by shifting the magnetic field $B_0$ or making the cavity resonant with the shifted cyclotron frequency. However, there is a residual effect on the cyclotron linewidth. This can be calculated by finding the standard deviation of the frequency shift $\Delta \omega_+/\omega_+$, $\sigma(\Delta \omega_+/\omega_+) = \frac{B_2}
{2B_0}\sqrt{\left<\bar{z}^4\right>-\left<\bar{z}^2\right>\left<\bar{z}^2\right>}$, which can be evaluated directly using the equations in Appendix \ref{Appendix:standardDeviation}, when $\bar{z}\gg \bar{z}_0$ 
\begin{align}
\sigma\left(\frac{\Delta \omega_+}{\omega_+}\right)&\simeq\frac{ B_2\bar{z}z_0}{\sqrt{2}B_0}.
    \label{eq:energyScatter2}
\end{align}
The evaluation of Eq.~\ref{eq:energyScatter2} is plotted in Fig.~\ref{figure:PhaseScatterAndAmplitudeNoise}~b). We showed in Section \ref{sec:4} that the electron-cavity linewidth in our region of interest was around the same as the bare cavity linewidth $\kappa$. This means that to avoid broadening the cyclotron resonance beyond this width, the presence of the $B_2$ term implies a maximum initial axial amplitude  
\begin{align}
    \bar{z}_i&\simeq\frac{B_0}{n Q B_2} \sqrt{\frac{2m \omega_z}{\hbar} }\, , \label{Eq:maxAmp}
\end{align}
where $n$ is the ratio between the cyclotron linewidth due to natural decay and the linewidth due to $B_2$ broadening. We normally set $n=2$ in the following discussion, which, as the linewidths combine in quadrature, leads to a 10\% reduction in absorption probability.   
\subsection{Including detector Johnson noise}
We now consider the phase uncertainty at the end of the photon counting sequence and incorporate the effect of the detector Johnson noise. The optimum detection time in this situation, considering the form of $\sigma_D$ from Eq.~\ref{eq:sigmaD}, is $\tau\simeq2.5/\gamma_z$. At this optimum, the noise associated with detection is 
\begin{align}
    \sigma _D^2&\simeq\frac{\frac{e_n^2}{4 |Z(\omega_z)|}+\frac{ k_B \textrm{Re}[Z(\omega_z)] T_D}{|Z(\omega_z)|}}{m \omega _z^2}.
\end{align}
A parallel LCR circuit will be used for the detection, in which case $\textrm{Re}[Z(\omega)]=R_p \mathcal{L}(\omega)$ and $|Z(\omega)|=R_p \sqrt{\mathcal{L}(\omega)}$ where $\mathcal{L}(\omega)$ is a Lorentzian function of unit height  
\begin{align}
\mathcal{L}(\omega)=\frac{1}{1+\frac{4Q_D^2(\omega-\omega_D)^2}{\omega_D^2}}.
\end{align}
Here $Q_D$ is the quality factor of the LCR circuit and $\omega_D$ is its resonant frequency. So, $\sigma _D^2$ becomes
\begin{align}
    \sigma _D^2&\simeq\frac{\frac{e_n^2}{4R_p\sqrt{\mathcal{L}(\omega)}}+ k_B T_D\sqrt{\mathcal{L}(\omega)}}{m \omega _z^2}.
\end{align}
If the particle oscillation frequency is close to $\omega_D$ the second term dominates, while if the particle is far from resonance, the first term is more significant. At the optimum detunings of the particle from the resonator,  
\begin{align}
    \omega &=\omega_D\pm \sqrt{\frac{\omega _D^2 \left(4 k_B T_D R_p-e_n^2\right)}{4 e_n^2Q_D^2}},
\end{align}
the noise reaches its minimum value and takes on the form 
\begin{align}
    \sigma _D^2&\simeq\frac{T_\textrm{eff}k_B}{m \omega _z^2},
    \label{eq:detection noise}
\end{align}
where $T_\textrm{eff}=\sqrt{\frac{e_n^2 T_D}{k_BR_P}}$. We have measured GaAs MESFET amplifiers with $e_n=0.7$ \si{\nano\volt\per \sqrt\Hz}, while copper coils can have $R_p\simeq 300$ k$\Omega$,   which makes $T_\textrm{eff}=0.35$ K at $T_D=1$ K. Put another way, an optimized detection process with these parameters adds an effective noise source to the axial amplitude equivalent to the thermal motion of a particle at 0.35 K. If the measurement had been conducted on-resonance, with $\omega_z$ matching $\omega_D$, the noise would be fixed at $\sigma_D^2 =\frac{T_D k_B}{m \omega_z^2}$, almost three times as high, and could only be further reduced by cooling the detection system. This is challenging as the detectors all dissipate some power and so cannot typically operate below 1 K. Detecting slightly off-resonance gives us more scope to reduce this noise contribution by increasing $R_p$ further using superconducting resonant coils or decreasing $e_n$ by switching from GaAs MESFETs to SQUIDs, in which case the benefits of measuring off-resonance would be more pronounced. 

While the foregoing analysis shows how the noise contribution of the detection system can be minimized, it still remains unacceptably high. To further suppress this noise, it is necessary to amplify the axial and magnetron modes to satisfy the condition
\begin{align}
    \sigma_{2,z}^2\gg\sigma_D^2.
\end{align} If we let $\sinh( kt_a)=27$, which amounts to an amplification of the axial amplitude $|\bar{z}_i|$ by a factor 27, then using $\sigma_D^2 = \frac{T_\textrm{eff} k_B}{m \omega_z^2}$ and $T_\textrm{eff}=0.35$ K, we have $\sigma_{z,2}^2=10\sigma_D^2$. The resulting amplified distribution corresponds closely to Eq.~\ref{eq:amped}, with $\sigma_{z,2}$ increased by 5\%. Since the amplification is large, $c^2(t)\simeq s^2(t)$. In this case the moments calculated from the amplified distribution become almost the same as those calculated from the distribution before amplification. The only difference is that $\sigma$ is increased by a factor of $\sqrt{2}\times1.05$, the factor $\sqrt{2}$ from the extra noise associated with the additional zero point energy from the second mode, and the factor $1.05$ from the residual influence of $\sigma_D$.

\begin{figure*}[t]
\centering
\includegraphics[]{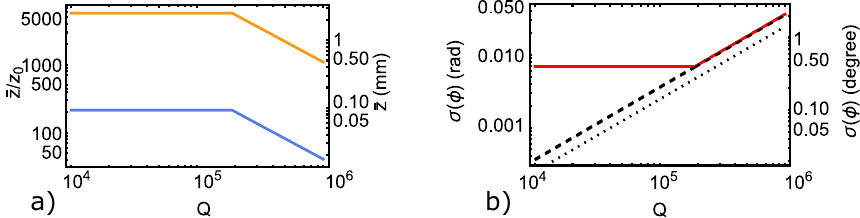}
\caption{a) Initial axial amplitude $|\bar{z}_i|$ from Eq.~\ref{eq:axAmp} in blue, and final axial amplitude $|\bar{z}_i||s(t_a)|$, assuming $|s(t_a)|=27$, in orange. b) $\sigma_\phi$ based on $|\bar{z}_i|$ from Eq.~\ref{eq:axAmp}, plotted in red. Also denoted by a dashed line is $\sigma_\phi$ if there were no limit on $\bar{z}_\textrm{max}$. A dotted line plots $\sigma_\phi$ for the hypothetical case where the detector noise is negligible and therefore no amplification is needed.}
\label{figure:realPhaseScatter}
\end{figure*}

The amplification pulse places an additional constraint on $\bar{z}_i$, namely that $|\bar{z}_f|=27|\bar{z}_i|<|\bar{z}_\textrm{max}|$, where $\bar{z}_\textrm{max}$ is the maximum feasible axial amplitude that can be used in the trap, limited by the trap geometry. Taking into account the constraints from both $B_2$ and the amplification, the initial axial amplitude can therefore be no greater than 
\begin{align}
    |\bar{z}_i|=\sqrt{2}z_0|\alpha_i|=\textrm{min}\left(\frac{\bar{z}_\textrm{max}}{|s(t_a)|},\frac{B_0}{n Q B_2} \sqrt{\frac{2m \omega_z}{\hbar} }\right)\, . 
    \label{eq:axAmp}
\end{align}
The associated phase uncertainty including the increased noise caused by amplification is then given, to a good approximation, by 
\begin{align}
  \sigma_\phi&\simeq\sqrt{\frac{2\hbar}{ m \omega_z\bar{z}_{i}^2}}.
  \label{eq:sigmaphi}
\end{align}
A plot of $|\bar{z}_i|$ and $|\bar{z}_i||s(t_a)|$ is shown in Fig.~\ref{figure:realPhaseScatter}~a) in blue and orange, respectively, as a function of the trap $Q$-factor. In this case, we have assumed $\bar{z}_\textrm{max}=2.5$~mm. At $Q$-factors lower than $Q=10^5$, where the cyclotron linewidth of the trap is comparatively broad, $|\bar{z}_i|$ is limited by the condition $\bar{z}_f<\bar{z}_\textrm{max}$. Above $Q=10^5$, the overriding limitation is that $\bar{z}_i$ should not lead to excess broadening by $B_2$, which places a more stringent limit on the initial amplitude. 

Fig.~\ref{figure:realPhaseScatter}~b) plots the phase uncertainty $\sigma_\phi$ found using the final axial amplitude after amplification, in red. This uncertainty shows a dependence on $Q$ that is the inverse of the blue trend shown in Fig.~\ref{figure:realPhaseScatter}~a). Also shown as a black dashed line is the phase uncertainty that would be achievable if there were no $\bar{z}_\textrm{max}$ limit. The dotted line  corresponds to $\sigma_\phi$ when $\sigma_D$ is so small that the amplification stage can be dispensed with, and there is no $\bar{z}_\textrm{max}$ limit. This dotted line is a factor of $\sqrt{2}$ lower than the dashed line. We note in passing that this figure also shows the important difference between the Pulse and Phase (PnP) \cite{Cornell1990ModeCrossing} and Pulse and Amplify (PnA) \cite{Sturm2011Phase-sensitiveEnergies} methods often used in Penning trap experiments. If detector noise $\sigma_D$ is negligible compared to the particle temperature noise, the phase noise achievable with PnP is a factor of $\sqrt{2}$ lower than PnA. The PnA method allows signals to be amplified to the level where  $\sigma_D$ can be neglected, at the cost of increasing the noise by a factor of $\sqrt{2}$, if both modes used in the PnA amplification stage are thermalized to the same average quantum state prior to the measurement.   

In this section, we have considered the phase noise we would expect in a simple realization of the photon counting experiment, and we have explored the limits imposed by $B_2$ and the need to amplify the signals above the detector Johnson noise. In the next section, we consider additional contributions to phase uncertainty imparted by further frequency shifts.  

\subsection{Frequency Shifts}

Electrostatic and magnetic imperfections, the magnetic bottle term $B_2$, and the relativistic mass increase all lead to frequency shifts which could potentially disrupt the photon detection method. We have already discussed how the $B_2$ term, in combination with axial amplitude noise, places limits on the initial excitation amplitude. In this section, we review the effect of other frequency shifts in detail. The form of the first-order shifts is described in Ref. \cite{Brown1986GeoniumTrap}, while higher-order shifts are summarized in e.g. Refs. \cite{Borchert2020, Major2005ChargedTraps, Ketter2014First-orderTrap}. The pertinent points to note are:

\begin{enumerate}
    \item The complete form of the electrostatic potential $V$ and the magnetic scalar potential $\Psi$ can be expanded in cylindrical coordinates $r$ and $\phi$ using Legendre polynomials $P_k(\cos(\phi))$  from the centre of the trap according to 
    \begin{align}
        V&=\frac{1}{2}V_0 \sum_{k=0}^{\infty}\left(\frac{r}{d}\right)^kC_k P_k(\cos(\phi)),\\
        \Psi&=- \sum_{k=1}^{\infty}k^{-1}B_{k-1}r^kP_k(\cos(\phi)),
    \end{align}
    where $d$ and $V_0$ have been defined previously. The coefficients $C_k$, and $B_k$, for $k>2$ are unwanted imperfections, while $C_2$ is the term responsible for axial trapping and $B_2$ is the deliberately introduced bottle.
    \item The $C_n$ coefficients are dimensionless and depend only on the trap geometry, while the $B_n$ coefficients have units of \si{\tesla\meter}$^{-n}$ and so depend on both the geometry and material properties through the magnetic field strength and magnetization. 
    \item The axial reflection symmetry means that only even powers need to be considered in these sums, so we neglect odd $k$ from now on.  
    \item In general, each term $C_k$ for $k>2$ and $B_{k'}$ for $k'>0$, leads to separate shifts of $\omega_+$, $\omega_z$ and $\omega_-$, where each shift depends on the total energy in the axial, magnetron and modified cyclotron modes. 
    \item Any increase in the particle energy in any mode leads to some increase in speed which causes a relativistic mass increase. This shift is most significant when the axial and modified cyclotron energy change, and has the biggest impact on the axial and modified cyclotron frequencies. 
\end{enumerate}

\begin{figure}[ht]
\centering
\includegraphics[width=0.4\textwidth]{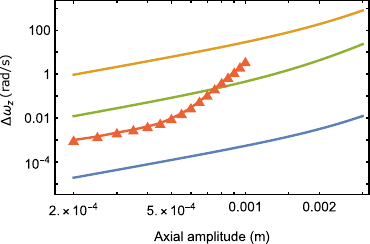}
\caption{Axial frequency shift due to electrostatic imperfections in a 3.7 mm radius 7 pole trap. Blue line represents an upper limit from an ideal trap, green a trap with $5$~\si{\micro\meter} manufacturing tolerances, and yellow $15$~\si{\micro\meter} tolerances. Red points are from Ref.\cite{FlorianThomasKohler2015Bound-ElectronIons}, scaled to an axial frequency of 100 \si{\mega\hertz}.}
\label{figure:axialFrequencyShifts}
\end{figure}

In an compensated 7-pole trap it should be possible to set all coefficients $C_k$ with $k>2$ to zero up to $k=10$ by controlling 5 trap voltages and grounding two end electrodes. In a realistic trap, manufacturing tolerances mean that the cancellation may not be perfect, and residual shifts from these terms may persist. Fig.~\ref{figure:axialFrequencyShifts} shows the typical range of axial frequency shifts which arise from electrostatic and relativistic perturbations in a 7-pole trap as the axial amplitude is varied. The blue line corresponds to our simulation of an optimized 7-pole design for a 3.7 mm radius trap; it is slightly non-zero because of the time limit placed on the optimization. The green and orange lines represent a trap with manufacturing tolerances of 5~\si{\micro\meter} and 15~\si{\micro\meter}, respectively. The red data correspond to an actual 7-pole trap \cite{FlorianThomasKohler2015Bound-ElectronIons}, scaled to an axial frequency of 100 \si{\mega\hertz}.  We use the data from Ref. \cite{FlorianThomasKohler2015Bound-ElectronIons} as a reasonable estimate for what can be achieved in a Penning trap. From these data, we see there is a residual shift $\Delta\omega_z=c_E|z|^2$,  $c_E=2.4\times10^4$ rad\,s$^{-1}$m$^{-2}$ expected in our trap. We can attribute this to a residual $C_4$ contribution $c_E=\frac{3 \omega _z C_4 }{4 d^2C_2}$. 

Similarly to the electrostatic imperfections, an ideal magnetic bottle produces a pure $B_2$ field, while a realistic bottle also produces an unwanted $B_4$ contribution and higher-order terms.  Our numerical calculations indicate that without much care, values of $B_4\approx 10^9 $~\si{\tesla\per\meter^4} are typical. With some careful design of the location of magnetic material and potentially some shimming, simulations show it should be possible to limit $B_4 < 10^7$ \si{\tesla\per\meter^4}. We now discuss the unwanted effects that $B_2$, $B_4$ and $C_4$ have on the detection process. Throughout, we are primarily concerned with the variance in the frequency shifts, which exists because of the axial and magnetron zero-point motion. This frequency noise can lead to phase noise in the final phase we wish to measure.  

\subsubsection{Shifts at the point of axial excitation}
\begin{figure*}[t]
    \centering
    \includegraphics[]{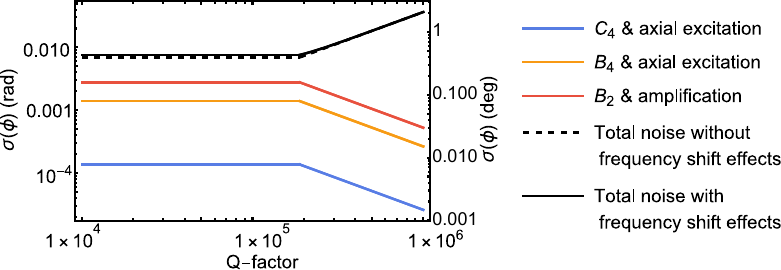}
    \caption{Various sources of phase noise, as a function of the cavity $Q$-factor, for the parameters listed in Appendix \ref{Appendix:trapParams}. The blue line shows the noise contribution from $C_4$ and the initial axial amplitude noise, the orange line shows the noise contribution from $B_4$ and the initial axial amplitude noise, the red line shows the noise contribution from $B_2$ and the axial and magnetron variance created during the amplification stage discussed in Section \ref{sec:fduringAmp}. Also plotted is the expected phase noise, ignoring the additional noise sources discussed in this section (black dashed) and including these noise sources (black solid).}
    \label{fig:FrequencyNoise}
\end{figure*}

The zero-point fluctuations in the axial motion in the ground state mean that, after excitation, there is an inevitable spread in the axial energy
\begin{align}
    \sigma(E_z)=\frac{\hbar\omega_z\bar{z}_i}{\sqrt{2}{z_0}}.
\end{align}
Inserting this energy variance into the expressions for the shifts caused by $c_E\propto C_4$ and $B_4$ (the $B_2$ term does not cause an axial energy-dependent axial frequency shift), we can determine the associated frequency noise using the parameters in Appendix \ref{Appendix:trapParams}. To relate this to a worst-case phase noise at the end of the counting sequence, we assume a maximum integration time of 100 s. The resulting phase noise is plotted in Fig.~\ref{fig:FrequencyNoise} in blue and orange. The phase noise arising from the axial zero point fluctuations and the finite size of $\bar{z}_i$ is also plotted as a black dashed line--this is the same as the red line in Fig.~\ref{figure:realPhaseScatter}~b).  It can be seen that the phase noise from the amplitude noise of initial axial excitation is far smaller than the phase scatter associated with the axial zero-point noise and can be neglected.  

\subsubsection{Frequency shifts during amplification, axial cooling and axial-magnetron \mathinhead{\pi}-flip}
\label{sec:fduringAmp}
The axial amplitude grows significantly during amplification, and the magnetron radius is boosted from its ground state. Both effects impart additional phase shifts onto the final phase detected after the $\pi$-flip. The electrostatic effects arise due to the variance in axial and magnetron amplitude, in combination with $C_4$, which leads to a variance in the magnetron frequency after excitation. As the axial cooling takes a finite time and there is a deliberate delay between the start of the axial cooling and the $\pi$-flip, this frequency variance translates into phase noise. Fortunately, for the parameters in Appendix \ref{Appendix:trapParams}, these effects lead to negligible additional phase noise. More troublesome is the equivalent magnetostatic effect, where the variance in axial and magnetron amplitudes after amplification combine with the large $B_2$ term to produce additional noise. The magnetron frequency shift after the amplification is given by \cite{Brown1986GeoniumTrap}
\begin{align}
    \Delta\omega_-&=\Delta\omega_{-,z}+\Delta\omega_{-,-}\\
    \Delta\omega_{-,z}&=-\frac{\omega _z^2B_2}{4\omega _+B_0}\bar{z}^2\\
    \Delta\omega_{-,-}&=\frac{\omega _z^2B_2}{4\omega _+B_0}\rho_-^2
\end{align}
Notice how an increase in the magnetron radius leads to a shift of the opposite sign to the shift when the axial amplitude increases. This is as expected, since the shift is driven by the quadrupolar $B_2$ term, which has opposite curvature in the axial and radial directions. We saw above that, after amplification, the axial and magnetron amplitudes are strongly correlated. We can use this correlation, in combination with a carefully chosen wait time, to almost completely cancel this phase shift and reduce its variance to an absolute minimum. 

We assume for simplicity that the amplification time $t_a$ is much shorter than the inverse of the energy damping constant $\gamma_\textrm{fast}$ in the detector's rapid cooling mode. In this case, the total magnetron phase shift imparted by the axial excitation after the axial amplitude has been completely cooled is  $\Delta\omega_{-,z}/\gamma_\textrm{fast}$. We choose the waiting time between the end of the parametric amplification pulse and the $\pi$-pulse to be equal to $t_\textrm{wait}=\frac{\omega _z}{2 \omega _-\gamma_\textrm{fast}}$, so that at the moment of the $\pi$-pulse the magnetron shift caused by the magnetron amplitude is $\Delta\omega_{-,-}t_\textrm{wait}$. The total magnetron phase shift after amplification due to the finite cooling time is then 
\begin{align}
\Delta \phi _-&=\frac{B_2 \omega _z^2}{4B_0 \gamma_\textrm{fast}   \omega _+}\left(\frac{\rho_- ^2}{\frac{2 \omega _-}{\omega _z}}-\bar{z}^2\right)
\end{align}
This can be rewritten in terms of the amplitudes at the end of the parametric amplification $\alpha_f$ and $\beta_f$
\begin{align}
\Delta \phi _-&=-\frac{\omega _z^2B_2}{2\gamma_\textrm{fast}  \omega _+B_0 }\frac{\hbar }{m \omega _z}\left(|\alpha_f|^2-|\beta_f|^2\right)
\end{align}
The standard deviation of this phase shift can be immediately calculated using Eq.~\ref{eq:amplitudeVariance}
\begin{align}
\sigma(\Delta \phi_-)&=-\frac{\hbar |\alpha'|\omega _zB_2}{2m\gamma_\textrm{fast}  \omega _+B_0  }
\end{align}
This noise is plotted in Fig.~\ref{fig:FrequencyNoise} in red. Although this is the largest shift, its size is manageable for the values of $|\bar{z}_i|$ we envision using, and it does not add a significant amount of extra noise.  

\subsubsection{Frequency shifts during detection}

During the detection period, the axial amplitude decreases as the particle is cooled by the detection system. If the axial frequency changes dramatically during the cooling period, it is more challenging for the required phase information to be recovered, and this may have an impact on the observed phase noise. In the discussion of the phase noise during the detection period, we considered only small frequency shifts during the detection. The most stringent requirement for this to be valid is that the frequency drifts by no more than one frequency bin of the Fourier transformed image current signal. As the optimum acquisition time is $\tau\simeq2.5/\gamma_z$, this means that the drifts in frequency should be no more than $\Delta\omega_z=\gamma_z/2.5$. Using $R_p = 300$ \si{\kilo\ohm} and $D_\textrm{eff,z}=12.75$ \si{\milli\meter}, we find $\gamma_z = 2\pi\times8.4$ \si{\hertz}. A typical final maximum axial amplitude after amplification will be around 2.5 mm. For the electrostatic imperfections, again using the measured trap data from Ref. \cite{FlorianThomasKohler2015Bound-ElectronIons}, this corresponds to a frequency shift of 1.8 Hz, well within the detection bandwidth $1/\tau$.

\subsection{Other experimental imperfections}
\subsubsection{Voltage noise}

Fluctuations $\sigma(\omega_z)$ in the trapping frequency $\omega_z$ during the axial free evolution period $t_\textrm{ev}$ due to voltage fluctuations can lead to phase noise $\sigma(\phi)=\sigma(\omega_z) t_\textrm{ev}$. For our parameters, we need the fractional voltage noise $\sigma\left(\frac{\Delta V_0}{V}\right)$ to be controlled better than $\sigma\left(\frac{\Delta V_0}{V}\right)<\textrm{max}(Q,10^5)\times1.6\times10^{-16}$ per second of evolution time $t_\textrm{ev}$. The best axial stability in a Penning trap has been achieved using three stacked Stahl electronics UM-14 supplies and a 2000 s time constant low pass filter \cite{Fan2022AnMoment}. The long-term drift in this experiment is $\sigma\left(\frac{\Delta V_0}{V}\right)=t\times5\times10^{-12}$, which would allow a free evolution time of around 1 second.  Higher voltage stability with shorter filter time constants can be achieved with a programmable Josephson voltage standard \cite{Kaiser2024JosephsonExperiments}, which should allow longer averaging times to be reached. A complementary solution is to connect the voltage source to a second trap with a second electron and use measurements in this auxiliary trap to correct for voltage drifts \cite{Pinegar2009StableExperiments}. 

\subsubsection{Heating rates}
Heating of the modified cyclotron, axial and magnetron modes needs to be carefully controlled in the experiment. The thermal microwave photons in the cavity are a background noise source which adds dark counts. The cavity occupation number is
\begin{align}
   \bar{n}&= \frac{1}{\exp \left(\frac{\hbar\omega_+ }{k_B T_e}\right)-1} \, .
\end{align}
For a critically coupled cavity, this cavity occupation number is equivalent to an incoming photon rate $R_a=\frac{1}{2}\kappa_c \bar{n}$. For $Q>10^4$, $T<70$ mK, and $\omega_+>2\pi\times 30$ GHz, the dark count rate is always less than $0.01$ counts/s. As we will see later, a count rate from axion conversion of $R_a>0.1$ counts/s is needed for a feasible axion search experiment, much larger than the dark count rate. Hence, cyclotron dark counts do not contribute meaningfully to the noise compared to other noise sources. 

The experiment is more robust with respect to axial or magnetron heating than it is to cyclotron heating. Although a full description of this is beyond the scope of this paper, we expect that any heating that occurs will add phase noise rather than destroying the measurement completely.  Starting with the axial mode, we note that, except when cooling is desired, the detection system is far detuned from the axial resonance. The requirement is that $\gamma_z\ll1/t_\textrm{ev}$, which can be achieved by switching a capacitance parallel to the LCR circuit. To estimate the axial heating rates, we can use the cyclotron heating rates achieved in other cryogenic Penning trap experiments and scale them to our trapping frequencies and particle mass. Ref. \cite{Borchert2019MeasurementTrap} demonstrated a cryogenic Penning trap with an electric field spectral density of $S_E\leq7.5_{-2.8}^{+3.4}\times10^{-20}$~\si{\volt\square\per\meter\square\per\hertz} at a frequency of $\omega_+=2\pi\times17.845$~\si{\mega\hertz}. This electric field density would lead to an axial transition rate out of the ground state for the electron at $\omega_z=2\pi\times100$~\si{\mega\hertz},
\begin{align}
    \zeta_z=\frac{q^2}{4m\hbar\omega_z}S_E,
\end{align}
of $\zeta_z\lesssim 0.01$~\si{\per\second}, which would allow evolution times up to 100 s before, on average, a photon absorption event occurs. We might also expect that $S_E$ might scale favorably in our proposed trap compared to Ref. \cite{Borchert2019MeasurementTrap}. The precise scaling of $S_E$ with trap dimensions $d$, frequency $\omega$ and temperature $T$ is complicated and shows large experiment-to-experiment variation \cite{Brownnutt2015Ion-trapSurfaces}. However, using a scaling $S_E\propto\omega^{-0.6}d^{-2}T^{1/2}$ which appears to be empirically justified for a cryogenic trap \cite{Brownnutt2015Ion-trapSurfaces}, we might expect the electric field density to be further reduced to $S_E\simeq1\times10^{-21}$~\si{\volt\square\per\meter\square\per\hertz}, which would lead to $\zeta_z\lesssim 0.6$ per hour, which is negligible. 

The magnetron mode, since it is at a much lower frequency, is more susceptible to noise. Repeating the transition rate calculation for this mode yields $\zeta_-\leq7$~\si{\per\second} for unscaled $S_E$ and $\zeta_-\leq4$~\si{\per\second} assuming $S_E$ varies as expected with frequency, temperature and distance. These rates would be somewhat limiting for the experiment's free evolution, so to match the limit imposed by voltage noise, further reduction of magnetron heating rates by at least an order of magnitude would be necessary. This could potentially be achieved by using notch filters tuned specifically to the magnetron frequency on the lines providing the trap voltages. 

\section{Calculating the detection efficiency}
\label{sec:6}
\subsection{Figure of merit}
We now consider the detection efficiency of the electron single-photon counter under various relevant scenarios. Before describing the results of more detailed calculations, we first present some simple arguments which illustrate the most important parameters that influence the detection efficiency. We start with the maximum initial excitation amplitude, before broadening effects occur, which is  given by Eq.~\ref{eq:axAmp} and is approximately, for large $Q$,  
\begin{align}
    \bar{z}_\textrm{max}&\simeq\frac{B_0}{n Q B_2} \sqrt{\frac{2m \omega_z}{\hbar} }\, . 
\end{align}
For large initial excitation amplitudes, and where the parametric amplification is sufficiently large that the thermal detection noise is negligible, the standard deviation of the final axial phase derived in Section \ref{sec:4} is approximately
\begin{align}
  \sigma_\phi&\simeq\frac{\hbar B_2n Q}{B_0 m \omega_z}  \,.
  \label{eq:approxScatter}
\end{align}
Detection will be most efficient when the phase advance $\Delta\phi$ as a result of a photon being present in the cavity is as large as possible compared to $\sigma_\phi$. On average, $\Delta\phi=\Delta\omega_z\langle t_{ex}\rangle$ where $\Delta\omega_z$ is given by Eq.~\ref{EqAxialShift} and $\langle t_{ex}\rangle$ is the average time the electron spends in the excited cyclotron state every time a photon is present in the cavity. The overall detection efficiency is then highest when the following figure of merit is maximized:
\begin{align}
    \frac{\Delta\phi}{\sigma_\phi}&=\frac{1}{n} \kappa \langle t_{\text{ex}}\rangle=\frac{1}{n} \, . 
\end{align}
The figure of merit is useful because it shows that provided the assumptions detailed above are valid, almost all the choices of trap parameters do not have a fundamental influence on the detection efficiency, and can be chosen to allow other unwanted frequency shifts to be manageable. Recall that $n$ is the ratio of the cavity absorption linewidth to magnetic broadening, and in the following sections we choose $n=2$. In practice, however, the figure of merit is only an approximation to the performance of the device; rather than the mean shift $\Delta\phi$, what is more significant is the value of the survival function $S(\phi)$ at $\phi\simeq\sigma_\phi$, which does depend on $g$ and $\kappa$.    

We now calculate the efficiency. We can use the phase measured after a photon counting sequence to test two different hypotheses, i) that a photon was present in the cavity and ii) that there was no photon in the cavity. This leads to two types of efficiency, which we call detection efficiency $\eta_d$ and exclusion efficiency $\eta_e$; these tell us respectively the efficiency with which we can detect a microwave single photon if it is present in the cavity, and the efficiency with which we can conclude that a microwave signal is not present, in both cases compared to an ideal single-photon counter.  

\subsection{Detection efficiency}

The detection efficiency $\eta_d$ is defined as the probability that a photon detection event is registered, if a photon is initially present in the cavity: $\eta_d=P(\textrm{Detection}|\textrm{Photon present in cavity})$. In a detector with readout noise, such as ours, defining the detection efficiency first requires establishing the significance level $\alpha_s$ with which we want to reject the hypothesis that the count was actually caused by a background process and not an incoming photon. In our electron single-photon counter, the axial phase is the single number reported at the end of a photon counting sequence, referenced to the average axial phase recorded when no photons are incident on the cavity. We assume that this reference phase is stable and known sufficiently well from preliminary measurements that it contributes no additional uncertainty. The distribution of axial phase differences from this mean, in the case that there are no incident photons, is a Gaussian $h_0(\phi)$, with zero mean and a standard deviation $\sigma_\phi$ given by Eq.~\ref{eq:sigmaphi}. 

\begin{figure}[ht]
\centering
\includegraphics[width=0.4\textwidth]{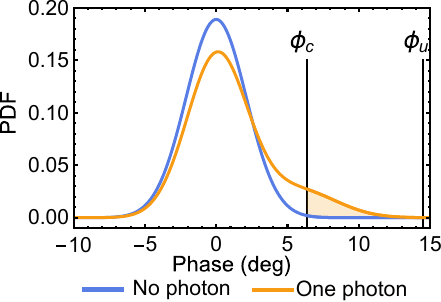}
\caption{An illustration of the portion of $h_1(\phi)$ (plotted in yellow) falling between $\phi_c$ and $\phi_u$ which is used to define the detection efficiency $\eta_d$. The false positive rate is found by integrating $h_0(\phi)$ (plotted in blue) between the same limits. For this plot $Q=10^6$ and $\sigma_c=3\sigma_\phi$, other parameters as in Tab. \ref{tab:placeholder}.}
\label{figure:photonAbsorbtion}
\end{figure}

If a photon is present in the cavity, the distribution of final phases $h_1(\phi)$ is given by convolving the Gaussian noise distribution $h_0(\phi)$ with $p_\phi(\phi)$, the probability density function for the axial phase advance found in Section \ref{sec:4},
\begin{align}
    h_1(\phi)&=(h_0*p_\phi)(\phi)=\int_{-\pi}^{\pi} h_0(\phi-\phi')p_\phi(\phi')d\phi' \,.
\end{align}
An example of the distributions $h_0(\phi)$ and $h_1(\phi)$ is shown in Fig.~\ref{figure:photonAbsorbtion}, here $h_1(\phi)$ is the PDF when one photon is present at some point in the cavity, and $h_0(\phi)$ when no photons are present. We define two phases $\phi_u$ and $\phi_{c}$, also shown in  Fig.~\ref{figure:photonAbsorbtion},  with $\phi_c\ll\phi_u\ll\pi$ such that if the measured phase $\phi$ satisfies $\phi_c<\phi<\phi_u$ after a photon counting sequence, we consider that a photon has been detected. The value of $\phi_c$ is found by requiring the probability of a false positive to be $\alpha_s$, which, considering that $h_0(\phi)$ is Gaussian distributed and so long as $\sigma_\phi\ll\pi$ is approximately given by 
\begin{align}
    \phi_c&=\sqrt{2} \sigma_\phi  \text{erf}^{-1}(1-2 \alpha_s) \, , 
\end{align}
where $\text{erf}^{-1}$ is the inverse error function. 
The detection probability $\eta_d$ is then simply given by
\begin{align}
    \eta_d=\int_{\phi_c}^{\phi_u} h_1(\phi)d\phi \, . 
\end{align}
An illustration of the integrated portion of $h_1(\phi)$ between $\phi_c$ and $\phi_{u}$ is shown in Fig.~\ref{figure:photonAbsorbtion}.

\begin{figure}[htb]
\centering
\includegraphics[]{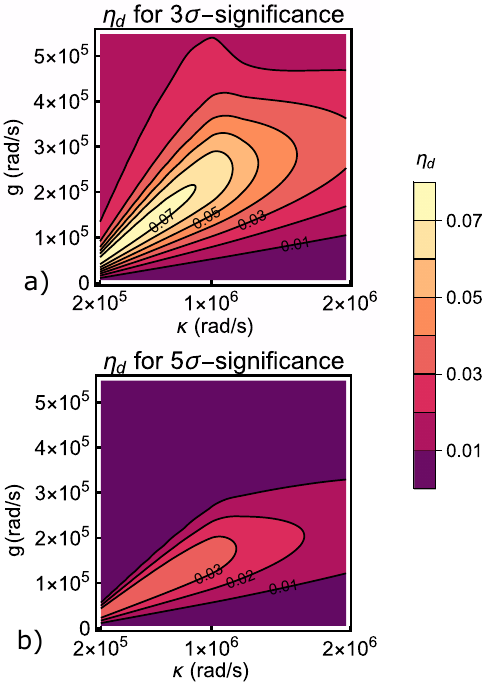}
\caption{Detection efficiency $\eta_d$ calculated from $h_1(\phi)$ as a function of $\kappa$ and $g$. a) is for 3-sigma significance and b) 5-sigma significance.}
\label{fig:gandkappaMap}
\end{figure}

In general, $p_\phi(\phi)$ is found by numerically integrating the equations for the cavity occupation as described in Section \ref{sec:4}. In the case that $g\ll\kappa$ it is possible to use an approximate form $p_\phi(\phi)\simeq\tilde{p}_\phi(\phi)$ valid for $\phi\geq0$ and hence provide an analytical expression for $\eta_d$. When $g\ll\kappa$, we model $p_\phi(\phi)$ as the sum of a delta function distribution centered at $\phi=0$ and a single-sided exponential distribution
\begin{align}
\tilde{p}_\phi(\phi) =
\begin{cases}
(1-P_e)\,\delta(\phi) + P_e \lambda e^{-\lambda \phi} & \phi \ge 0, \\[0.5em]
0 & \phi < 0,
\end{cases}
\end{align}
with $\lambda=\frac{4 g^2}{\kappa}\frac{ m \omega _z B_0}{\hbar \omega _+ B_2}$, and under the assumption  $\lambda^{-1}\ll\pi$ as is the case for the situations we consider. This probability distribution has a simple physical interpretation: in the weak coupling limit $g\ll\kappa$, a photon is either absorbed by the electron, with probability $P_e$ or is not absorbed, with probability $1-P_e$. If there is no absorption, then no additional phase is imparted, while if the absorption occurs, the excited state lifetime follows an exponential decay with a time constant  $\tau=\frac{\kappa}{4 g^2}$. Relating the time in the excited state to the phase accrued leads to the expression for $\lambda$. The convolution of $\tilde{p}_\phi(\phi)$ with $h_0(\phi)$ can also be computed analytically, with the result
\begin{align}
    \tilde{h}_1(\phi)=&(1-P_e)\frac{e^{-\frac{\phi^2}{2 \sigma_\phi ^2}}}{\sqrt{2 \pi } \sigma_\phi }\nonumber\\&+\frac{1}{2} \lambda  P_e e^{\frac{1}{2} \lambda  \left(\lambda  \sigma_\phi^2-2 \phi \right)} \left(\text{erf}\left(\frac{\phi -\lambda  \sigma_\phi ^2}{\sqrt{2} \sigma_\phi }\right)+1\right).
    \label{eq:happrox}
\end{align}
This approximate form allows for a rapid calculation of $\eta_d$ in the weak coupling case, but breaks down when $g\simeq\kappa$. 
\begin{figure}[htb]
\centering
\includegraphics[width=0.43\textwidth]{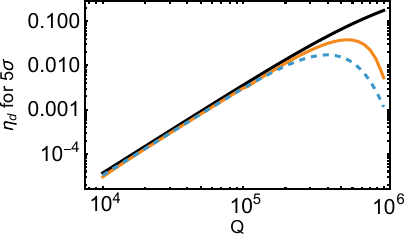}
\caption{The detection efficiency for a 5-sigma CL, plotted in orange. Also shown in black is the maximum possible detection efficiency given by Eq.~\ref{eq:maximum}. The blue dashed lines show the detection efficiency calculated from the analytical approximation to the phase distribution function $\tilde{h}_1(\phi)$.}
\label{figure:detectionEfficiency}
\end{figure}
The results of the full calculation of $h_1(\phi)$ and the approximation $\tilde{h}_1(\phi)$ are plotted in Fig.~\ref{figure:detectionEfficiency}. Both these efficiencies have been multiplied by $(1+n^{-2})^{-1/2}=\sqrt{4/5}$ to account for the loss in efficiency due to $B_2$ induced broadening. Here Fig.~\ref{figure:detectionEfficiency}~a) shows $\eta_d$ requiring a 5-sigma significance for detection, ($\alpha_s\simeq2.9\times10^{-7}$). The dashed blue line represents $\tilde{h}_1(\phi)$, while the solid orange line shows the full form of $h_1(\phi)$ without approximation. Also shown in black is $P_e$, which represents the theoretical maximum efficiency if every photon absorption led to a phase advance in the photon counter of sufficient size that it could unambiguously be attributed to an absorption event. The performance of the counter tracks this line at low $Q$, but deviates at higher $Q$. This is the region where the axial phase noise leads to a loss in detection efficiency. The decrease in efficiency once $Q$ passes an optimum value can be attributed to the reduction in excited state lifetime due to enhanced spontaneous emission rate, a lower excited state lifetime outweighing the benefit of a higher absorption probability. Fig. \ref{fig:gandkappaMap} plots $\eta_d$ calculated using $h_1(\phi)$ for 3- and 5-sigma significance as a function of $\kappa$ and $g$, where the plotted $\eta_d$ is again multiplied by $\sqrt{4/5}$. These plots show that the efficiency peaks in a diagonal region at around $g=\kappa/4$ for the 3-sigma confidence and $g=\kappa/6$ for 5-sigma confidence at lower values of $g$ and $\kappa$, but then declines at higher values of both parameters. This decline is caused by the lower limit on $\sigma_\phi$ set by the finite trap size as plotted in Fig. \ref{fig:FrequencyNoise}. At the optimium values of $g$ and $\kappa$, the maximum detection efficiency is $\eta_d=7.4\%$ for 3-sigma and  $\eta_d=3.8\%$ for 5-sigma confidence respectively. 

\subsection{Exclusion efficiency}
We now turn to the exclusion efficiency $\eta_e$, which we define as $\eta_e(C_L)=\frac{t_\textrm{real}(C_L)}{t_\textrm{ideal}(C_L)}$. Here $t_\textrm{ideal}(C_L)$ is the waiting time required for an ideal single-photon counter to exclude an incident photon rate $R_a$ with a certain confidence level $C_L$ and $t_\textrm{real}(C_L)$ is the waiting time for the electron photon counter we have been considering. The inverse $\eta_e^{-1}$ tells us, on average, how much longer we would have to wait to exclude a certain count rate using the electron single-photon counter, compared to an ideal device. 

For low photon count rates such as we expect in axion experiments, the number of photons appearing in an interval $t$ should conform to a Poisson distribution. For an ideal photon counter, the wait time required can be found from the probability of zero counts being observed after a wait time $t$, $e^{-R_at}=1-C_L$. Typically, a $95\%$ exclusion interval $C_L=0.95$ is required, so that $t_\textrm{ideal}\simeq3/R_a$. In order to find $\eta_e$ we must find the equivalent time $t_\textrm{real}(C_L)$ for our photon counter. We imagine that $N$ photons come sequentially, and the phase readout occurs after all of them have arrived. We find the distribution of phases $h_N(\phi)$ after these $N$ photons by performing repeated convolutions of the phase PDF $p_\phi(\phi)$ and the noise PDF $h_0(\phi)$ 
\begin{align}
    h_N(\phi)&=(h_0*\underbrace{p_\phi*...*p_\phi}_N)(\phi) \, . 
    \label{eq:exclusionConvolution}
\end{align}
The average number of photons $N=R_at_\textrm{real}$ required to exclude the rate $R_a$ at confidence level $C_L$ is found by calculating the function
\begin{align}
    I(N)=1-\int_{\phi_l}^{\phi_u} h_N(\phi)d\phi \, ,
\end{align}
and then interpolating the results to find the $N$ such that $I(N)=C_L$. The phases $\phi_l$ and $\phi_u$ can be chosen to maximize $C_L$ for a given $N$. The interpolated value of $N$ (not necessarily an integer) which satisfies this equation then defines  $t_\textrm{real}=N/R_a$. Using the phase that falls between $\phi_l$ and $\phi_u$ rather than the phase larger than a single bound allows us to manage the situation where the phase wraps around from $\pi$ to $-\pi$. In order to associate a single efficiency $\eta_e$ with the process, we consider that the measured phase falls within $\phi_l$ and $\phi_u$, which are typically close to $\phi=0$. This means that, with the chosen wait time, we would expect to exclude the given rate $R_a$ limit with confidence $C_L$ or higher, in the following proportion of experimental runs where the measured phase falls between $\phi_l$ and $\phi_u$
\begin{align}
\frac{1}{\sqrt{2 \pi } \sigma _{\phi }}\int_{\phi_l}^{\phi_u}e^{-\frac{\phi^2}{2 \sigma_\phi} }d\phi .    
\end{align}
\begin{figure}[htb]
\centering
\includegraphics[width=0.4\textwidth]{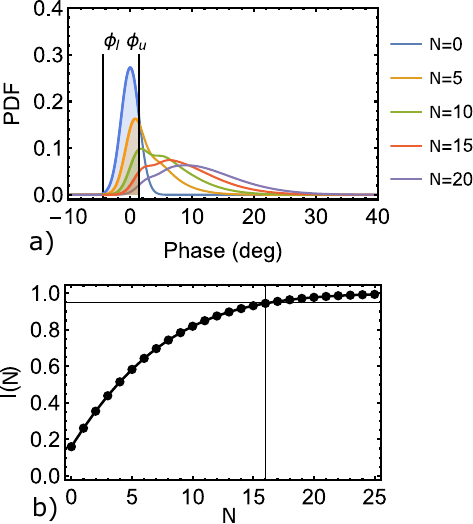}
\caption{The procedure to define the exclusion efficiency $\eta_e$. a) $h_N(\phi)$ is plotted for several values of $N$. The segment of $h_N(\phi)$ between $\phi_l$ and $\phi_u$ is shown with color filled to the $x$-axis. b) Shows $I(N)$, and the point at which $I(N)=0.95$ is marked by lines.}
\label{figure:photonExclusion}
\end{figure}
Fig.~\ref{figure:photonExclusion} illustrates the two stages of this calculation. Fig.~\ref{figure:photonExclusion}~a) shows $h_N(\phi)$ for $N=\{0,5,10,15,20\}$. As $N$ increases, 
$h_N(\phi)$ is shifted to higher values and less of the distribution falls between $\phi_l$ and $\phi_u$. Fig.~\ref{figure:photonExclusion}~b) shows $I(N)$ as $N$ increases. The point where $I(N)=0.95$ is also indicated.    

If the phase is outside the interval between $\phi_l$ and $\phi_u$, then $R_a$ can either be excluded at lower confidence or a higher $R_a$ can be excluded at the same confidence. 
As with consideration of $\eta_d$, an analytical integral $\tilde{h}_N(\phi)$ found by evaluating Eq.~\ref{eq:exclusionConvolution} with $\tilde{p}_\phi(\phi)$ instead of $p_\phi(\phi)$ can be derived, and it is expected to be valid in the limit that $g\ll\kappa$ and provided that effects related to the wrapping around of the axial phase are not significant. The expression for $\tilde{h}_N$ is provided in Appendix \ref{Appendix:erlang}. 

Fig.~\ref{figure:photonExclusionEfficiency} shows the 95\% exclusion efficiency calculated for the electron single-photon counter, setting $\phi_u=\sigma_\phi$, $\phi_l=-3\sigma_\phi$, with other parameters listed in the caption. This choice of $\phi_u$ and $\phi_l$ means that at least 84\% of measurements would expect to set a 95\% confidence limit or better on a given photon rate. The rates derived from $\tilde{h}_N(\phi)$ and $h_N(\phi)$ (dashed blue and orange) have been multiplied by $0.84(1+n^{-2})^{-1/2}=0.84\sqrt{4/5}$ to account for the loss in efficiency due to $B_2$ induced broadening and the fact that this efficiency is only achieved on 84\% of measurements. 

\begin{figure}[htb]
\centering
\includegraphics[width=0.4\textwidth]{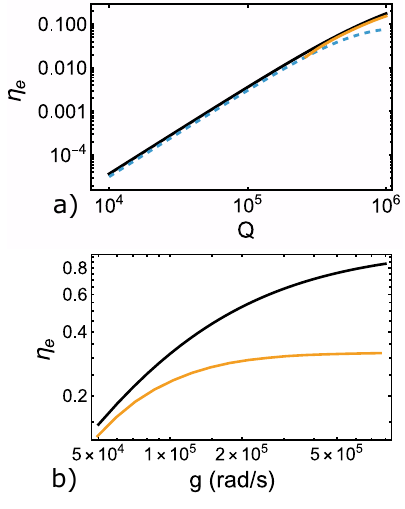}
\caption{Exclusion efficiency $\eta_e$ for $\phi_l=-3\sigma_\phi$, $\phi_u=\sigma_\phi$. Figure a) plots the efficiency as a function of $Q$ and $\omega_+=2\pi\times30$ GHz, b) as a function of $g$ for $Q=10^6$ and $\omega_+=2\pi\times30$ GHz. The orange lines plot the full calculation using $h_N(\phi)$. The blue dashed line gives the approximation derived from $\tilde{h}_N(\phi)$, and the black line is the maximum possible efficiency using the absorption probability from Eq.~\ref{figure:detectionEfficiency}. The orange and blue efficiencies have been multiplied by $0.84\sqrt{4/5}$ as discussed in the text.}
\label{figure:photonExclusionEfficiency}
\end{figure}

Comparing Fig.~\ref{figure:photonExclusionEfficiency}~a) and Fig.~\ref{figure:detectionEfficiency}, we see that the exclusion efficiency is generally higher than the detection efficiency, and that the optimum $Q$-factor for the most efficient detection is higher. This is because phase shifts from individual photon absorptions, which are individually too small to be resolved against the phase noise, can nevertheless add together and lead to a measurable shift when $R_at$ is large. Fig. ~\ref{figure:photonExclusionEfficiency}~b) plots $\eta_e$ as a function of $g$ when $Q=10^6$  (equivalently $\kappa=188,496)$. This shows that the exclusion efficiency continues to rise as with increasing $g$, and saturates at about $\eta_e=0.30$ when $g$ and $\kappa$ become comparable. 



\section{Discussion and Conclusions}
\label{sec:7}
Here we discuss the implications of what we have discovered for axion detection, and compare the expected performance of our scheme to other proposals for photon counting with trapped electrons. 

Our development of a single-photon counter is motivated by the aim of speeding up axion search experiments. In the introduction, we noted that for axion experiments where $Q_c\approx Q_a$, at 30 GHz, and assuming $\eta_\textrm{la}=1$ a single-photon counter with an efficiency of $\eta_\textrm{sp}>3\times10^{-5}R_a$ will outperform a linear amplifier. Assuming axions solve the strong CP problem and make up all of dark matter, we could expect a haloscope to produce of order 0.001-1 counts/s at 30 GHz depending on the design. We can consider five different scenarios for the electron-photon counter depending on the electron trap construction materials and design. The corresponding $Q$-factors, coupling strength $g$, exclusion efficiencies and speedup times are shown in Table \ref{tab:photonCounting}, assuming an additional $\eta_c=0.5$ coupling loss between the amplifier and the axion-to-photon conversion device of the type introduced in Section \ref{sec:2}, so that the total exclusion efficiency is $\eta_t=\eta_c\eta_e$.

\begin{table}[htb]
\begin{center}

\begin{tabular}{|l|c|c|c|c|c|}
\hline
Cavity type                    & $Q$  & \begin{tabular}[c]{@{}c@{}}$g$\\ (krad/s) \\ \end{tabular}            & $\eta_e (\%)$ & \begin{tabular}[c]{@{}c@{}}Speedup vs. \\  linear amplifier \\ \end{tabular} \\ \hline
i) Copper        & $2\times 10^4$ &$2\pi\times 9.1$   &   $0.01$       &  1.4 \\ 
ii) NbSn$_3$          & $6\times 10^4$ &  $2\pi\times 9.1$ &    $0.08$       &   13 \\ 
iii) Copper/NbTi          & $4\times 10^5$ & $2\pi\times 9.1$  &    $4.2$       &   630  \\ 
iv) Thin NbTi & $1\times10^6$    & $2\pi\times 9.1$ &   $15$  &  2300 \\ 
v) Thin NbTi & $1\times10^6$    & $2\pi\times 30$    & $29$      &  4300 \\ \hline
\end{tabular}
\caption{Parameters for various potential photon counters. For the last column, $R_a=1$~counts/s, $\eta_{la}=1$, $Q_c\simeq Q_a$, $\eta_c=0.5$. }
\label{tab:photonCounting}
\end{center}
\end{table}

We see that even a modestly efficient single-photon counter speeds up the axion search, while the ultimate device leads to a 4300-fold increase for a 1 Hz axion count rate.

\begin{figure}[ht]
\centering
\includegraphics[]{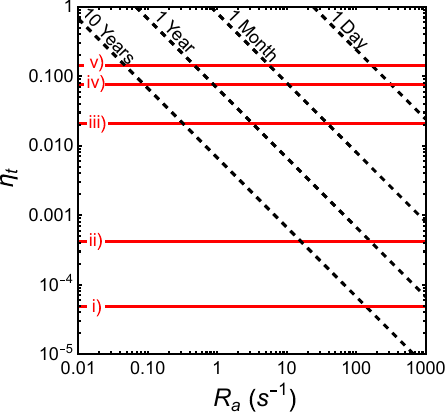}
\caption{Total integration time required to exclude an axion signal at 95\% confidence over a frequency octave $m_a\rightarrow2m_a$ as a function of $R_a$ and $\eta_t$, indicated by dashed lines. The red horizontal lines indicate the photon counting efficiencies envisioned in Tab.~\ref{tab:photonCounting}.}
\label{fig:optPlot}
\end{figure}

Another way in which the usefulness of the photon counter can be put in context is by considering the total integration time required to scan a particular range of axion masses, given that the axion haloscope produces a count rate $R_a$. Assuming once again that $Q_c\simeq Q_a$ and $\eta_c=0.5$, the time required to scan between an axion mass $m_a$ and $2 m_a$ assuming that the central frequency is increased by a factor $1+1/Q_a$ each time, is 
\begin{align}
t_s=\frac{3 \log (2)}{R_a\eta _t\log \left(\frac{1}{Q_a}+1\right)}.
\end{align}
 The dashed lines in Fig.~\ref{fig:optPlot} correspond to $t_s$ for one day, one month, one year, and ten years, as a function of  $\eta _t$ and $R_a$. Also shown on the plot are horizontal lines representing the five photon counting scenarios considered in Tab. \ref{tab:photonCounting}. We consider that an experiment which takes 10 years of measurement time to measure an octave from $m_a$ to $2m_a$ is the limit of feasibility. We see from this figure that scenario i) will only be feasible with an axion haloscope that produces count rates in excess of $R_a=140$~s$^{-1}$. On the other hand, scenario ii) can allow experiments with count rates $R_a>16$~s$^{-1}$, while iii), iv) and v) allow experiments with count rates $R_a>0.32$~s$^{-1}$, $R_a>0.08$~s$^{-1}$ and $R_a>0.05$~s$^{-1}$ respectively. The count rates for scenarios iii)-v) could be achieved with novel haloscope designs.

Our ultimate goal is a Penning trap with $Q=10^6$, and with a $g$ controllable between $g\simeq2\pi\times6\times 10^3$ (for high detection efficiency) and $g>2\pi\times3\times 10^4$ (for high exclusion efficiency). A variable $g$ could potentially be achieved by moving the electron position within the electric field of the cavity. Alternatively, we could vary the number electrons in the trap and work with the collective axial mode of small electron clouds, which boosts $g$ by the square root of the number of electrons. 

We now compare our method to other proposals for counting single photons using trapped electrons. Cridland et al., \cite{Cridland2016SingleElectron} propose trapping a single electron above a surface Penning trap that doubles as a microwave coplanar waveguide. The basics of the method are similar to ours: a trap with a $B_2$ is used to cause the electron axial frequency to shift on absorption of the photon. The electron-waveguide mode coupling is analyzed using transmission line theory. To boost the coupling between the trapped electron and the microwave photon traveling in the waveguide, the electron is trapped at a position where the admittance of the waveguide (the reciprocal of the impedance) is as small as possible compared to the effective admittance of the electron, which is typically a few tens of pS. It is proposed that low waveguide impedance can be achieved by a shorted stub microwave termination an electrical distance corresponding to $\lambda/4$ away from the electron position. It is estimated that dielectric losses are due to the sapphire substrate of the coplanar waveguide at cryogenic temperatures, and an admittance of 900 pS can be achieved, which gives a few \% absorption efficiency if the electron is around 100 \si{\micro\meter} above the trap surface. This is similar to the mid-range efficiencies achievable with our method. The method of Ref. \cite{Cridland2016SingleElectron} has the principal advantage that, as it uses a waveguide rather than a cavity, the sensitive frequency of the device can be adjusted by simply adjusting the magnetic field of the trap, rather than, in our case, also having to move the endcaps to keep the cavity resonant with the modified cyclotron mode. The surface trap method also has technical challenges, chief among these being maintaining a low admittance at the electron position. We note that a small displacement away from the optimized position can dramatically increase the admittance, which reduces the absorption efficiency. Excited axial amplitudes are needed for the types of phase-sensitive detection methods proposed in Ref. \cite{Cridland2016SingleElectron}, which may be challenging to reconcile with the narrow range of permissible electron positions. Trapping electrons above surface traps is challenging, as yet single electrons have not been observed in such traps due to complications arising from the non-ideal trapping potential. If these challenges are overcome, we note that the method of Ref. \cite{Cridland2016SingleElectron} could be combined with our approach by using a coplanar waveguide cavity rather than a transmission line. This structure would have a much higher coupling constant $g$ than a cylindrical trap, enabling the $g=\kappa$ strong coupling regime to be reached at a lower $Q$-factor. 

A more recent proposal by Fan et al., \cite{Fan2025HighlyDetection} has a different approach on how to boost the electron-photon interaction. In this proposal, the orbital electric dipole moment of the electron is increased by placing the particle into a highly excited cyclotron state $n_+=10^6\left(\frac{\omega_a}{0.1~\textrm{meV}}\right)^2$. The electron then behaves similarly to a Rydberg atom, with an enhanced interaction to electric fields, with a coupling rate that scales like $n_+$. As in our proposal and Ref. \cite{Cridland2016SingleElectron}, the cyclotron state in Ref. \cite{Fan2025HighlyDetection} is detected using a $B_2$ bottle and by measuring the axial frequency. The signature in the Fan et al., proposal is a one-quantum increase in the cyclotron state, superimposed over the decaying cyclotron signal. Comparing the efficiency of this method to our proposal is complicated, as Ref. \cite{Fan2025HighlyDetection} reports detection limits for a complete axion detector that includes both the single electron photon counter and a large magnetized mirror like the BREAD antenna \cite{PhysRevLett.128.131801}. We observe that using a state with $n_+=10^6$ in a Penning trap of radius $r=0.5$ mm and length $l=1.0$ mm as envisioned in Ref. \cite{Fan2025HighlyDetection} would achieve $g\simeq\kappa$ with $Q\sim1$, a regime which happens in our cavity at $Q=10^6$. This means that we expect the overall absorption probability to be similar between our two devices, neglecting broadening effects. Ultimately, it is an open question whether it will be easier experimentally, as we propose, to keep the electron close to the ground state and look for small phase advances due to photon absorption, or attempt to measure far larger advances in a highly excited state. 

In this work, we have described a new technique for counting single photons using an electron in a Penning trap. The method is optimized for use in axion searches, where a photon counter can speed up experiments operating above 30 GHz by several orders of magnitude. The new method will enable axions searches above $\sim100$~\si{\micro\electronvolt} in a reasonable experimental time. In future work we will describe how this electron single-photon counter can be combined with high $Q$-factor axion conversion cavities under development in our laboratory to make an axion haloscope able to probe DFSZ axions above 120 \si{\micro\electronvolt}.
\newpage
\clearpage
\onecolumngrid
\begin{appendices}
\section{}

\label{Appendix:trapParams}
The typical experimental parameters used in this paper are as follows
\begin{table}[ht]
    \centering
    \begin{tabular}{| c|c|c|c|}
    \hline
    Parameter name  & Symbol &Value & Unit \\
    \hline
      Modified cyclotron frequency   & $\frac{\omega_+}{2\pi}$& $30$ & \si{\giga\hertz} \\
      Axial frequency   & $\frac{\omega_z}{2\pi}$&$100$ & \si{\mega\hertz} \\
      Magnetron frequency   & $\frac{\omega_-}{2\pi}$&$167$ & \si{\kilo\hertz} \\
        Axial frequency shift from $C_4$   & $c_E$ & $2.4\times10^4$ & \si{\radian \,m^{-2}} \\
      Static magnetic field   & $B_0$ & $1.07$ & \si{\tesla} \\
      Magnetic bottle strength   & $B_2$ & $10^5$ & \si{\tesla\per\meter\squared} \\
    Quartic magnetic field correction  & $B_4$ & $<10^7$ & \si{\tesla \meter^{-4}} \\
      Trap length & $l$ & 24.55 & \si{\milli\meter}\\
      Trap radius & $r$ & 3.7 & \si{\milli\meter}\\
      Trap electric volume & $\tilde{V}$ & $2.4\times10^{-7}$ & \si{\meter\cubed} \\
      Cavity frequency   & $\frac{\omega}{2\pi}$& $30$ & \si{\giga\hertz} \\   
      Cavity coupling constant & $\frac{g}{2\pi}$ & 9.2  & \si{\kilo\hertz} \\
      Free phase evolution time & $t_\textrm{ev}$ & 1-100 & \si{\second}\\
      Detection system parallel resistance & $R_p$ & 300 & \si{\kilo\ohm}\\
      Detection system effective electrode distance & $D_\textrm{eff,z}$ & 12.75 & \si{\milli\meter}\\
      $B_2$ broadening to cavity linewidth ratio & $n$ & 2 & None \\
      Maximum axial amplitude & $z_\textrm{max}$ & 2.5& mm \\
      Axial amplification factor & $\sinh kt_a$ & 27& None \\
      Detector noise & $\sigma_D^2$ & $\frac{0.35 k_B}{m \omega_z^2}$ & \si{\meter\squared}\\
      \hline
    \end{tabular}
    \caption{Trap parameters used to evaluate expressions throughout this paper, unless otherwise stated.}
    \label{tab:placeholder}
\end{table}
\FloatBarrier
\section{}
\label{Appendix:cyclotronEmission}
In this appendix, we confirm that the matrix elements (\ref{eqnMatrixElement1}-\ref{eqnMatrixElement2}) are consistent with the familiar expression for the rate of cyclotron emission of an electron in a magnetic field \cite{Jackson1998ClassicalElectodynamics, Brown1986GeoniumTrap}, 
\begin{align}
    \frac{d E}{d t }=-\frac{1}{4\pi\varepsilon _0}\frac{4e^2\omega ^2}{ 3m c^3}E \,.
\end{align}
Fermi's Golden Rule gives the transition rate from an initial state to a set of final states $f$ 
\begin{align}
    \Gamma_{i\rightarrow f}&= \frac{2\pi}{\hbar}|\langle i | H_\textrm{int}| f \rangle|^2 \rho(E_f)
\end{align}
where $\rho(E_f)$ is the density of final states corresponding to the final state energy $E_f$. To calculate the free space emission we need to consider that the photon can be emitted in any direction, so we need to integrate over all angles to get
\begin{widetext}

\begin{align}
   \Gamma_{i\rightarrow f}=& \frac{2\pi}{\hbar}e^2\frac{\hbar }{\varepsilon _0\omega \tilde{V}}\left(\frac{\omega _+}{\omega _+-\omega_-}\right)^2\rho(E_f)\nonumber\\&\times\int_0^\pi\int_0^{2\pi}|\langle i | \frac{1}{4\pi} \left(a_\gamma e^{i \boldsymbol{k}\cdot \boldsymbol{r}} + a_\gamma^{\dagger }e^{-i \boldsymbol{k}\cdot \boldsymbol{r}}\right)\cos(\theta)\boldsymbol{\hat{x}}\cdot\boldsymbol{V}^+ | f \rangle|^2 \sin(\theta)d\theta d\phi \,. 
    \label{gamma1}
\end{align}

\end{widetext}
where $\theta$ is the azimuthal angle and $\phi$ the polar angle, and the $x$-axis is chosen to lie along the projection of the emitted photon's direction onto the radial plane. The result is  
\begin{align}
   \Gamma_{i\rightarrow f}=& \frac{2\pi}{3\hbar}\rho(E_f)e^2\frac{\hbar }{\varepsilon _0\omega \tilde{V}}\left(\frac{\omega _+}{\omega _+-\omega_-}\right)^2\nonumber\\&\times|\langle i' | \left(a_\gamma e^{i \boldsymbol{k}\cdot \boldsymbol{r}} + a_\gamma^{\dagger }e^{-i \boldsymbol{k}\cdot \boldsymbol{r}}\right)\boldsymbol{\hat{x}}\cdot\boldsymbol{V}^+ | f' \rangle|^2 \,,
    \label{gamma2}
\end{align}
where the prime index indicates that the angular dependence of the final states has been removed. 

Now consider a decay from $n_+$ to $n_+-1$ where the cavity begins empty $n_\gamma=0$. As a given cyclotron mode can decay by emitting photons of two orthogonal polarizations via couplings between $\boldsymbol{\hat{x}}\cdot\boldsymbol{V}^+$ as considered so far and $\boldsymbol{\hat{y}}\cdot\boldsymbol{V}^+$, up to now ignored, the total decay rate for the cyclotron mode needs to be multiplied by a factor of 2. The terms to the right of the density of states $\rho(E_f)$ can be replaced using the matrix element Eq.~\ref{eqnMatrixElement2}, yielding
\begin{align}
   \Gamma_{n_+\rightarrow (n_{+}-1)}&= \frac{4\pi}{3\hbar}\rho(E_f)e^2\hbar^2\frac{n_+}{4 m\varepsilon _0\tilde{V}}
    \label{gamma3}
\end{align}
The density of final states, considering that there are two polarizations, is 
\begin{align}
    \rho(E_f)=\frac{\omega ^2 V}{\pi ^2 c^3 \hbar} \, ,
\end{align}
inserting this into Eq.~\ref{gamma3} and assuming that as the physical volume gets large, the mode volume $\tilde{V}\rightarrow V$, we arrive at the final expression for the transition rate out of the $n_+$ exited state as
\begin{align}
   \Gamma_{n_+\rightarrow (n_{+}-1)}&= \frac{1}{4\pi\varepsilon _0}\frac{4e^2\omega ^2}{ 3m c^3}n_+
    \label{gamma4}
\end{align}
As $n_+\propto E$ this is equivalent to the classical result we wanted to demonstrate. 
\section{}
\label{Appendix:quantumClassical}

In this appendix we show the equivalence of the quantum and classical rates for a charged particle interacting with the cavity in the weak coupling limit. We showed previously that the rate at which a particle in the excited cyclotron state decays and hence loses energy is given by 
\begin{align}
    \gamma_\textrm{q}=\frac{4 g^2}{\kappa}=\frac{q^2 Q}{m \tilde{V} \omega  \varepsilon_0}.
\end{align}

The classical rate at which a charged particle loses energy due to interacting with nearby conductors is given by the expression 

\begin{align}
    \gamma_\textrm{c}=\frac{q^2 R}{m D_{\text{eff, +}}^2} 
    \label{eqGammaC}
\end{align}

Here $R$ is the resistance experienced by the oscillating particle current as a result of nearby conductors and $D_{\text{eff, +}}$ is the effective electrode distance for the cyclotron mode, which also satisfies the relation
\begin{align}
    D_{\text{eff, +}}^2=\frac{V_{\text{induced}}^2}{E_p^2}
\end{align}

Here $E_p$ is the electric field at the position of the particle produced by the image currents in the conductors and $V_{\text{induced}}$ is the induced voltage on those conductors. 

We recognize that the power loss can be written in two different ways, first in terms of the induced voltage
\begin{align}
    P=\frac{V_{\text{induced}}^2}{R} .
\end{align}

And second in terms of the quality factor 
\begin{align}
    P={\omega U}{Q} .
\end{align}

Here $U=\int{\varepsilon_0|E|^2 dV}$ is the total power stored in the cavity system, with associated quality factor $Q$. Putting this all together with Eq.~\ref{eqGammaC} we have 

\begin{align}
    \gamma_\textrm{c}=\frac{q^2 Q}{m\omega  \varepsilon_0}\frac{E_p^2}{\int{|E|^2 dV}} 
    \label{eq:GammaC}
\end{align}

Which, given the previous definition for $\tilde{V}$ shows that $\gamma_\textrm{c}=\gamma_q$. Note that as discussed in the text, this equivalence no longer holds when $\kappa$ becomes comparable or smaller than $g$ and hence the classical expression for the radiated power should expect to break down when strong coupling occurs. 

\section{}
\label{Appendix:standardDeviation}
The probability distributions for the axial position after excitation and at the point of detection both have the form
\begin{align}
    \mathcal{F}(r,\theta)=\frac{1}{2\pi\sigma^2}\exp\left[-\frac{\left(r\cos(\theta)-a\right)^2+\left(r\sin(\theta)-b\right)^2}{2\sigma^2}\right]
\end{align}
where $a^2+b^2=c^2$ and $\langle X\rangle=\int_0^\infty\int_{-\pi}^\pi X r \mathcal{F} dr d\theta$. Here we provide some evaluations used throughout the paper
\begin{align}
    \langle r^2\rangle&=c^2+2 \sigma ^2\simeq c^2\, \label{eq:rsquared} ,\\
    \langle r^4\rangle-\langle r^2\rangle^2&=4\sigma ^2 \left(c^2+\sigma ^2\right)\simeq 4\sigma^2c^2\, .
\end{align}
where the approximation holds if $c^2\gg\sigma^2$. In this case, it is also true that
\begin{align}
    \langle \theta^4\rangle-\langle \theta^2\rangle^2&=\frac{2\sigma^2}{c^2}\, .
\end{align}

\section{}
\label{Appendix:erlang}
The repeated convolution of $\tilde{p}_\phi$ has a simple form in terms of an Erlang distribution

\begin{align}
  (\underbrace{\tilde{p}_\phi*...*\tilde{p}_\phi}_n)(\phi) \, &=\frac{\lambda^n \phi^{n-1} e^{-\lambda\phi}}{\Gamma (n)} \, ,
    \label{eq:exclusionConvolution2}
\end{align}
where $\Gamma (n)$ is the Euler gamma function. A further convolution with a normal distribution gives, via the Mathematica computer algebra package,  $\tilde{h}_n$ as
\begin{widetext}

\begin{align}
\tilde{h}_n(\phi)=&\frac{2^{\frac{n}{2}-2} \lambda ^n \sigma_{\phi }^{n-1} e^{-\frac{\phi^2}{2 \sigma _{\phi }^2}}}{\sqrt{\pi } \Gamma (n)} \Bigg[\sqrt{2} \Gamma \left(\frac{n}{2}\right) \, _1F_1\left(\frac{n}{2};\frac{1}{2};\frac{1}{2} \left( \lambda  \sigma _{\phi }-\frac{\phi}{\sigma _{\phi }}\right)^2\right)\\
&-2 \Gamma \left(\frac{n+1}{2}\right)\left( \lambda  \sigma _{\phi }-\frac{\phi}{\sigma _{\phi }}\right) \, _1F_1\left(\frac{n+1}{2};\frac{3}{2};\frac{1}{2} \left( \lambda  \sigma _{\phi }-\frac{\phi}{\sigma _{\phi }}\right)^2\right)\Bigg] \, .
\end{align}

\end{widetext}
Here $_1F_1(a;b;z)$ is the Kummer confluent hypergeometric function.

\end{appendices}
\bibliography{references2}
\end{document}